\documentclass[a4paper,11pt]{article}
\PassOptionsToPackage{table}{xcolor}
\usepackage{jinstpub} 
\usepackage{comment}
\usepackage{subcaption}
\usepackage{tabularx}
\usepackage{booktabs}
\usepackage{longtable}
\usepackage{tikz}
\usepackage{graphicx}
\usepackage{gensymb}
\usepackage{hyperref}
\usepackage{soul}
\usepackage{lineno}
\newcolumntype{C}{>{\centering\arraybackslash}X}

\title{\boldmath{DC Cryogenic Modeling of Open-Source SkyWater 130 nm MOSFETs at 77 K Using BSIM4}}
\author{F.~Beall$^1$, A.~Rimal$^1$, O.~Seidel$^1$, Y.~Mei$^1$, A.~D.~McDonald$^3$, I. Parmaksiz$^{5,1}$, V.~A.~Chirayath$^1$, J.~Asaadi$^1$, D.~Braga$^2$, J.~B.~R.~Battat$^4$}
\affiliation{$^1$The University of Texas at Arlington,\\
Physics Department, Arlington, TX 76019, USA}
\affiliation{$^2$Fermi National Accelerator Laboratory,\\
Microelectronics Department, Batavia, IL 60510, USA}
\affiliation{$^3$Instrumentation Frontier Scientific, Arlington, TX 76019, USA}
\affiliation{$^4$Wellesley College, \\ Physics and Astronomy Department, Wellesley, MA 02481, USA}
\affiliation{$^5$Rice University, \\Physics Department, Houston, TX 77005, USA}
\emailAdd{feb9528@mavs.uta.edu}

\abstract{
Cryogenic applications in high-energy physics (HEP) demand reliable, low-power CMOS electronics capable of operating at liquid nitrogen temperatures (77\,K). The open-source SkyWater 130nm (SKY130) CMOS process has previously been shown to operate at temperatures as low as 4\,K making it a promising candidate for HEP applications. In this work, we characterize and model SKY130 low-threshold voltage transistors at 77 K, which is a temperature commonly used in modeling applications for liquid argon detectors. DC characteristic measurements were performed at both room temperature and liquid nitrogen temperature. We created a cryogenic modeling approach to produce a SPICE-compatible, isothermal BSIM4-based model for select transistor sizes at 77\,K. The resulting model agrees with data at 77\,K with an average error on the order of $20\%$ (relative RMS). However, our models had a systematic overprediction specifically for the measured pMOS devices at the lowest measured drain bias.  Due to the open-source nature of SKY130, we have made our models publicly available on Github. We hope this work will continue the trend for democratizing circuit design at cryogenic temperatures in high-energy physics by enabling open access to accurate CMOS device models at 77\,K. 
}

\begin{document}

\maketitle

\section{Introduction}
Open-source process design kits (PDKs), which are open-source libraries made available by the semiconductor foundry that model the fabrication process and the variations there-in, offer a compelling alternative to the traditional high-cost, proprietary PDKs that restrict accessibility in the field of circuit design. Google has supported and collaborated on two open-source PDK efforts: SkyWater 130nm (SKY130) and GlobalFoundries 180nm \cite{GooglePDKs}. The Leibniz Institute for High Performance Microelectronics (IHP) released their own open-source BiCMOS IHP 130nm PDK \cite{IHP}.
Due to rising interest in cryogenic PDKs for cryogenic transistor operation, it is only natural for cryogenic PDKs to be developed from open-source processes. There was an effort to characterize and model SKY130 at 4\,K, demonstrating its operation at this temperature \cite{akturk2023cryogenicsky1304k}. Those behind the IHP SG13G2 130-nm BiCMOS open PDK have been making strides to develop a cryogenic PDK as low as 4\,K for integrated circuit design \cite{IHPcryo}. 

High-energy physics (HEP) experiments would benefit from open-source PDKs, which can lower barriers to access practical mixed-signal ASIC design on open technology platforms such as SKY130 \cite{akturk2023cryogenicsky1304k, Fath2024, raheem2025designADCsky130,souzadesignlinearOTA}. Many current and future HEP experiments rely on large-scale detectors based on liquid argon time projection chambers (LArTPCs) \cite{Acerbi2024,Asaadi2024,MicroBooNE:2016pwy,instruments8030041}. There are multiple advantages offered by placing the complementary metal-oxide-semiconductor (CMOS)-based ASICs for signal readout inside the cryostat. The first is that it significantly reduces noise as a result of lower thermal fluctuations. Placing the CMOS front-end ASICs in the cryostat and close to the electrode reduces cable length and input capacitance, which lowers the equivalent noise charge of LArTPCs. In addition, it decouples the cryostat and electrode design from the readout electronics, which simplifies the overall integration of the system \cite{Berns2019,gao2022cold}. Consequently, the ASICs operation must be understood and reliable within the 77-89K temperature range. Various CMOS technology nodes, from 180nm down to 28nm, have been characterized and modeled at 77\,K for this purpose \cite{cryo28QNTMBeckers2017,ZHAO201449,LUO201912,de2010front180nm,hoff2015cryogenic65nm130nm}. 

The SkyWater 130nm node, developed by SkyWater technology in collaboration with Google, is the first open-source CMOS technology available to the public \cite{skyDoc}. Previous studies have been performed to utilize this technology node for designing various systems, such as an ADC for control applications and a linear operational transconductance amplifier (OTA) for analog design at 300\,K ~\cite{Fath2024, raheem2025designADCsky130,souzadesignlinearOTA}, and a two-stage amplifier circuit at 4\,K~\cite{akturk2023cryogenicsky1304k}. 
However, the performance of SKY130 at liquid nitrogen and liquid argon temperatures (77--89\,K), the region of interest for LArTPCs, remains unexplored. Neither room temperature nor 4\,K isothermal models accurately predict transistor behavior at this intermediate temperature, due to the temperature dependent nature of various transistor parameters. 
The present study is restricted to low-threshold-voltage (LVT) nMOS and pMOS devices in the SKY130 process, motivated by their intended use in cryogenic front-end electronics for LArTPCs. Other threshold-voltage device options are outside the scope of this work. The measured geometries and associated calibrated model bins are listed in Tables {\ref{tab:pmos}} and {\ref{tab:nmos}}. To our knowledge, this work presents the first systematic DC characterization and modeling of the LVT SKY130 CMOS technology node at 77\,K.

In this work, we provide a brief overview of relevant CMOS physical parameters at cryogenic temperature and their corresponding BSIM4 parameters (Section~\ref{sec:cryoCMOS}). The cryogenic measurement setup used to acquire the data necessary for modeling is outlined in Section \ref{sec:experiment}. Experimental results at 77\,K are presented in Section~\ref{sec:measRes}. In Section~\ref{sec:cryomodeling}, we take an intensive look at the cryogenic modeling methodology and the resulting models which follow, also reporting on the accuracy of the produced 77\,K models. Section~\ref{sec:conclusion} concludes with possible future work in this node as well as a reference to the developed open-source 77\,K models.

\section{Cryogenic CMOS Behavior} \label{sec:cryoCMOS}
Effective CMOS design requires an understanding of transistor parameters and their impact on SPICE-level design. As temperature decreases, semiconductor device behavior changes accordingly, altering key transistor operating parameters. A number of these behaviors are represented in model parameters stored in the PDK, specifically the Berkeley Short-channel IGFET Model framework (BSIM4) model framework. BSIM4 is a compact, analytical, physics-based predictive MOSFET SPICE model for CMOS development and circuit simulation, accurate into the sub-100nm regime \cite{bsim4manual} where short-channel effects, such as drain-induced barrier lowering (DIBL), have significant impact on drain-current behavior. Such effects are captured in the BSIM4 model.
\newline

\subsection{BSIM4 Model Parameters at 77\,K}
The SKY130 PDK utilizes the BSIM4 model which contains a collection of parameters that are specific to a particular process node and whose values are verified by semiconductor foundries. Process parameters are related to transistor manufacturing, whereas compact (or basic) model parameters are used for circuit simulation. Many compact parameters model physical behaviors of the MOSFET under certain operating conditions and capture the temperature-dependent performance of the devices. The cryogenic transistor effects most relevant to this work and the corresponding BSIM4 parameters targeted in the 77\,K extraction are discussed below.

\subsubsection{Threshold Voltage ($V_{TH}$)} \label{vth_section}
Previous studies show that the MOSFET threshold voltage generally increases as temperature decreases \cite{cryoCMOSbeckers2018,VthcryoBeckers2020,VthmodelDao2017}. A major contribution to this trend is the temperature dependence of the bulk Fermi potential, because the intrinsic carrier concentration $n_i$ decreases strongly at low temperature. At cryogenic temperatures, however, the threshold voltage shift cannot be described only by the conventional complete-ionization form \cite{cryoCMOSbeckers2018,VthcryoBeckers2020,VthmodelDao2017}. Incomplete dopant ionization (freeze-out) changes the bulk Fermi level and the physical definition of inversion, making it important for accurate cryogenic $V_{TH}$ modeling even though it is not the sole origin of the overall temperature trend. Bandgap widening produces an additional upward shift in $V_{TH}$. Some compact models also include field-assisted dopant ionization as a correction term, whereas physics-based analyzes indicate that this process is largely completed before inversion and can often be neglected at threshold in a first-order treatment. In addition, interface traps, particularly traps near the band edge, can modify the low-temperature behavior and help explain deviations from simple saturation, especially in some pMOS measurements \cite{VthcryoBeckers2020}. Since $V_{TH}$ determines the turn-on condition of the MOSFET, these cryogenic shifts directly affect device biasing and the power/performance trade-offs in circuit design.

A detailed description of the $V_{TH}$ model used in BSIM4 can be found in Chapter 2 of Ref.~\cite{bsim4manual}. In this model one of the parameters in the expression for $V_{TH}$ is \verb=VTH0= which is defined as the long-channel threshold voltage at zero substrate bias. In our parameter extraction methodology, described in Section \ref{sec:ModelAndDesignParameters}, \verb=VTH0= serves as the reference parameter that mainly captures the measured shift in $V_{TH}$ between 300\,K and 77\,K.  

\subsubsection{ Carrier Mobility ($\mu$)} \label{mobility_section}

A key transport parameter in cryogenic MOSFET operation is the inversion-layer carrier mobility, which directly affects the drain current, transconductance, and on-state performance of nMOS and pMOS devices. At low temperature, the mobility cannot be described reliably by a simple room-temperature extrapolation. Instead, it is better understood as the result of competing scattering mechanisms in the inversion layer. Following the low-temperature semi-empirical treatment as given in Ref.~\cite{AroraGildenblat1987}, the effective mobility may be written using Matthiessen's rule as
\begin{equation}
    \frac{1}{\mu_{\mathrm{eff}}}
    =
    \frac{1}{\mu_{\mathrm{ph}}}
    +
    \frac{1}{\mu_{\mathrm{C}}}
    +
    \frac{1}{\mu_{\mathrm{sr}}},
\end{equation}
where $\mu_{\mathrm{ph}}$, $\mu_{\mathrm{C}}$, and $\mu_{\mathrm{sr}}$ represent the phonon-, Coulomb-, and surface-roughness-limited mobility components, respectively. As the temperature is reduced, phonon scattering weakens and tends to increase $\mu_{\mathrm{eff}}$. At the same time, however, Coulomb scattering from ionized impurities and interface charge, together with surface-roughness scattering at high vertical field, becomes relatively more important. As a result, cryogenic mobility reflects a balance among these mechanisms rather than a single monotonic temperature trend.

A useful way to parameterize this behavior is through the effective transverse electric field,
\begin{equation}
    E_{\mathrm{eff}}=\frac{Q_{\mathrm{dpl}}+\eta Q_{\mathrm{inv}}}{\varepsilon_{\mathrm{si}}},
\end{equation}
where $Q_{\mathrm{dpl}}$ and $Q_{\mathrm{inv}}$ are the depletion and inversion charge densities, respectively, $\varepsilon_{\mathrm{si}}$ is the permittivity of Si and $\eta$ is an empirical field-partition parameter. Investigations \cite{ChengWoo1997}  have shown that $\eta$ is itself temperature dependent and that the mobility-field characteristics of electrons and holes increasingly differ as the temperature is lowered. Therefore, although cryogenic operation often enhances the mobility of the inversion-layer, the resulting device behavior must be interpreted together with concurrent changes in the threshold voltage and subthreshold characteristics, which are discussed in the following section.

In BSIM4, the effective mobility ($\mu_{\mathrm{eff}}$) is modeled through a unified mobility formulation described in Chapter 5.2 of Ref.~\cite{bsim4manual}, with the low-field mobility parameter \verb=U0= setting the baseline mobility level. In the present work, we retain the mobility-model form used in the room-temperature PDK (Equation 5.6 in Ref.~\cite{bsim4manual}) and adjust \verb=U0= to capture the first-order change in effective mobility at 77\,K, while preserving the original field-dependence structure of the model.

\subsubsection{Subthreshold swing ($SS$)} \label{SS_section}
The transistor behavior in the subthreshold operating region is described by the subthreshold swing ($SS$) parameter. This parameter measures how much gate voltage is required to change the drain current by one decade, and is therefore commonly written as \cite{cryoSSBeckers2020}
\begin{equation}
    SS \equiv \frac{\partial V_{GS}}{\partial \log_{10}(I_D)}.
\end{equation}
A smaller $SS$ corresponds to a sharper turn-on characteristic and improved switching efficiency.

In the subthreshold region, the drain current is dominated primarily by diffusion current. Under the classical Boltzmann picture, the subthreshold swing is given by
\begin{equation}
    SS = n\frac{k_B T}{q}\ln(10),
    \label{SS_eq}
\end{equation}
where $k_B$ is the Boltzmann constant, $T$ is the device temperature, $q$ is the elementary charge, and
\begin{equation}
    n = 1 + \frac{C_{\mathrm{dep}}}{C_{\mathrm{ox}}} + \frac{C_{\mathrm{dsc}} + C_{\mathrm{it}}}{C_{\mathrm{ox}}}
\end{equation}
is the subthreshold swing parameter, with $C_{\mathrm{dep}}$, $C_{\mathrm{it}}$, $C_{\mathrm{ds}}$, and $C_{\mathrm{ox}}$ denoting the capacitance of the depletion width, the capacitance due to the interface states, the coupling capacitance between drain/source to channel, and oxide capacitances, respectively. The ideal Boltzmann limit corresponds to $n=1$, for which $SS=(k_B T/q)\ln (10)$ \cite{sub4KKamgar1982, cryoSSBeckers2020,28BulkBeckers2018}.

From room temperature down to liquid-nitrogen temperature, this thermal scaling provides a reasonable first-order description of experimental behavior. However, at deep-cryogenic temperatures (below 50\,K) the measured $SS$ deviates from the ideal thermal limit and eventually saturates due to non-ideal electrostatics and band-tail broadening. Since the present work focuses on operation near 77\,K, the classical thermal trend with a non-ideal slope factor remains the more relevant starting point \cite{sub4KKamgar1982}.

Because $SS$ determines how much gate-voltage change is required to switch the device between low-current and high-current operation, it is an important parameter for low-voltage and low-power circuit design. A smaller $SS$ generally improves switching efficiency and helps reduce the gate-voltage swing needed to achieve a given current ratio, although the overall off-state leakage also depends on other parameters such as threshold voltage.

In BSIM4, the subthreshold behavior is captured through  $n$ in the drain-current expression as explained in Chapter 3 of Ref.~\cite{bsim4manual}. Specifically, $n$ is rewritten with \verb=NFACTOR= in front of $\frac{C_{\mathrm{dep}}}{C_{\mathrm{ox}}}$  and is introduced to compensate for errors in the depletion-width capacitance calculation. In the present temperature-specific extraction methodology, the measured shift in subthreshold swing between 300\,K and 77\,K is captured primarily through \verb=NFACTOR=, described in Section \ref{sec:ModelAndDesignParameters}, with the remaining subthreshold parameters kept fixed \cite{bsim4manual}.

\subsubsection{On-State Current ($I_{on}$)} \label{Ion_section}
The on-state current $I_{on}$ is the drain current evaluated at a specified ON-state bias. In digital characterization, this bias is often chosen near $V_{GS}=V_{DS}=V_{DD}$, so $I_{on}$ is commonly close to a saturation-region current. From room temperature to cryogenic operation, $I_{on}$ is influenced by two competing trends: the increase in threshold voltage (Section~\ref{vth_section}) and the increase in carrier mobility (Section~\ref{mobility_section}). In the 28~nm bulk CMOS study of Beckers \textit{et al.}, the threshold voltage shifts upward while the low-field mobility increases strongly at low temperature; as a result, the on-state current at 4.2~K increases only for the long devices \cite{28BulkBeckers2018}.

In BSIM4, $I_{on}$ is obtained from the unified drain-current model described in Chapter~5 of Ref.~\cite{bsim4manual}. Under high lateral electric field, the onset of current saturation is influenced by velocity saturation, which BSIM4 models through the carrier-velocity relation
\begin{align*} 
    v &= \frac{\mu_{\mathrm{eff}}E}{1+E/E_{\mathrm{sat}}}
    && \text{for } E < E_{\mathrm{sat}} \\
    v &= \texttt{VSAT}
    && \text{for } E \ge E_{\mathrm{sat}}
\end{align*}
with $E_{\mathrm{sat}}$ satisfying the relation
\begin{equation}
    E_{\mathrm{sat}}=\frac{2\,\texttt{VSAT}}{\mu_{\mathrm{eff}}}.
\end{equation}
Here $v$ is the carrier velocity and $E_{\mathrm{sat}}$ is the critical electrical field at which $v$  becomes saturated. Thus, the parameter \verb=VSAT= influences the model-predicted onset of saturation (via the computed $V_{dsat}$) and the drain current in the high-$V_{DS}$ regime, and therefore impacts the extracted $I_{on}$ \cite{bsim4manual}. To ensure a smooth and numerically robust transition between the linear (triode) and saturation regions in the output characteristics, BSIM4 introduces an effective drain bias $V_{dseff}$, defined as a smooth function of $V_{DS}$ and $V_{dsat}$ \cite{bsim4manual}. The sharpness of the transition near the knee is controlled by the model parameter \verb=DELTA=, which shapes the $I_D$--$V_{DS}$ behavior in the vicinity of $V_{dsat}$ and helps maintain continuity of derivatives for circuit simulation \cite{bsim4manual}.  In the present 77~K extraction methodology, the measured change in saturation behavior is captured partly through \verb=VSAT=, together with the temperature-dependent shifts in threshold voltage and mobility \cite{bsim4manual}. \verb=DELTA= is modified if the data shows a non-ideal fit in the linear-to-saturation transition region of the output characteristics. 

\subsubsection{Drain-to-Source Series Resistance ($R_{sd}$)}\label{Rsd_section}

The source/drain series resistance of a MOSFET contains several contributions. For conventionally doped source/drain regions, Nguyen-Duc \textit{et al.} decompose the total resistance as
\begin{equation}
    R_{sd}=R_c+R_j+R_{sp},
\end{equation}
where $R_c$ is the contact resistance between the metal and the doped semiconductor, $R_j$ is the diffusion sheet resistance of the source/drain regions, and $R_{sp}$ is the spreading (injection) resistance associated with current crowding near the channel/source-drain interface \cite{RsdNguyenDuc1986}. In that case, the contact and diffusion contributions were found to decrease as temperature is lowered, and the overall impact of $R_{sd}$ becomes more significant as the channel length is reduced \cite{RsdNguyenDuc1986}.

For transistors with lightly doped drain (LDD) extensions, an additional cryogenic mechanism must be considered. LDD structures are introduced to reduce the electric field near the drain and thereby mitigate hot-carrier effects \cite{LDDAndhare1990}. However, at cryogenic temperatures the lightly doped regions may undergo impurity freeze-out. Hafez \textit{et al.} \cite{LDDcryoHafez1995} showed that below approximately 100\,K, the LDD resistance can increase strongly with decreasing temperature, substantially degrading the low-drain-voltage characteristics. Additionally, Hafez \textit{et al.} \cite{LDDcryoHafez1995} showed that at sufficiently high drain and gate voltages the LDD resistance becomes field dependent and decreases due to field-assisted impurity ionization. Thus, near cryogenic temperatures, the total series resistance can reflect two opposite trends: a decrease of the heavily doped source/drain parasitic resistance with decreasing temperature, and an increase of the lightly doped extension resistance due to freeze-out.

In BSIM4, the source/drain resistance is modeled in Section 5.3 through the bias-dependent source/drain resistance model \cite{bsim4manual}. For the internal symmetric resistance formulation (\verb=RDSMOD= = 0, Equation 5.17 in Section 5.3 of \cite{bsim4manual}), the relevant parameter is \verb=RDSW=, which BSIM4 defines as the zero-bias LDD resistance per unit width. In the present 77 K extraction methodology, \verb=RDSW= serves as the primary parameter used to capture the change in the overall source/drain series resistance \cite{bsim4manual}.

\subsubsection{Drain-Induced Barrier Lowering} \label{dibl_section}
Drain-induced barrier lowering (DIBL) is a short-channel effect in which an increase in drain bias lowers the source-channel potential barrier, leading to an apparent reduction in threshold voltage as $V_{DS}$ increases \cite{bsim4manual}. In BSIM4, DIBL is included in the threshold-voltage model of Chapter 2. Specifically, Equation~(2.32) defines the channel-length-dependent DIBL coefficient $\theta_{th}^{\mathrm{DIBL}}$, and Equation~(2.33) gives the corresponding threshold-voltage shift as
\begin{equation}
    \Delta V_{th}^{(\mathrm{DIBL})}
    = -\theta_{th}^{\mathrm{DIBL}}\left(\verb!ETA0! + \verb!ETAB! *\,V_{BS}\right)V_{DS}.
\end{equation}
Thus, within the BSIM4 framework, \verb=ETA0= is the primary DIBL coefficient in the subthreshold region, while \verb=ETAB= is the body-bias coefficient for the subthreshold DIBL effect \cite{bsim4manual}. In the present extraction methodology, when the 77\,K transfer characteristics showed visible DIBL, \verb=ETA0= was used as the primary fitting parameter to capture the drain-bias-induced threshold shift.

\section{Experimental Setup} \label{sec:experiment}

\begin{figure}[htbp]
\centering
\includegraphics[width=1\linewidth]{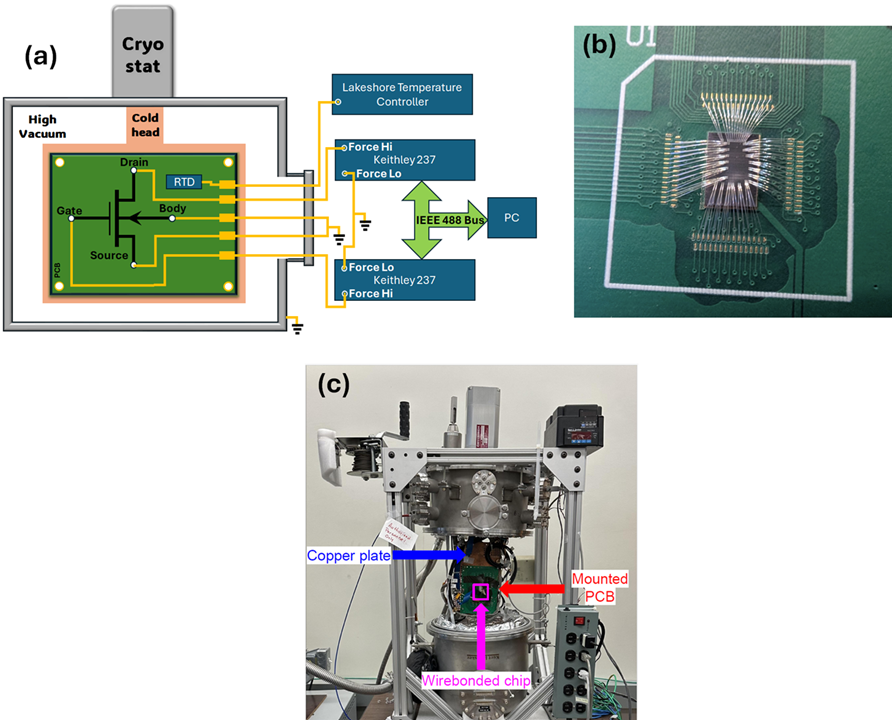}
\caption{Hardware setup used for I–V measurements: (a) Schematic of the I–V measurement system (b) Wirebonded SKY130 chip mounted on PCB, (c) Chip mounted on copper plate for 77\,K measurements.}
\label{fig:hardware_setup}

\end{figure}

The SKY130 MOSFETs characterized in this work were manufactured as a part of the Efabless Open Multi-Project Wafer (MPW) program \cite{efabless}. The fabricated devices include low-threshold voltage  transistors (LVT) and custom designed circuits like operational amplifiers, Ring oscillators, and integrators. The test-transistor set comprises 120 nMOS devices with gate lengths ranging from $0.15$ to $100~\mu\mathrm{m}$ and 50 pMOS devices with gate lengths
ranging from $0.35$ to $20~\mu\mathrm{m}$. Devices with channel lengths greater than $10~\mu\mathrm{m}$ are allowed by the PDK and were achieved by stacking transistors. While this is not a common use case, these long lengths were included in fabrication for a general understanding of device characteristics at large geometries. Two long nMOS were tested and included in the cryo-model (see Table \ref{tab:nmos}). For both polarities, the fabricated gate widths range from $0.42$ to $100~\mu\mathrm{m}$. The gate lengths and widths were selected with reference to the device geometries documented in the SKY130 PDK and in the latest NGSpice PDK patch \cite{skywaterpdk,skypdkgit}. Devices were selected for wire bonding with the aim of covering the available geometry options for both polarities. However,
some selected devices did not provide an electrical response
during testing, and measurements were therefore not obtained for every intended geometry. Tables {\ref{tab:pmos}} and {\ref{tab:nmos}}  list the pMOS and nMOS geometries, respectively, for which measurements were obtained and used in this study, rather than the complete set of geometries fabricated on the chip.

Current-voltage (I–V) characteristics were measured at 77\,K for twenty-two MOSFETs at Fermi National Accelerator Laboratory (FNAL). The devices were cooled to 77\,K using a single-stage cryo-cooler developed at FNAL. The test chip was wire-bonded onto a custom-designed PCB and covered with a 3D-printed plastic cap to protect the wire bonds, while maintaining stable access to all required pads. The PCB assembly was mounted on a copper plate inside a cryogenic vacuum chamber and cooled to 77\,K using a single-stage cryocooler. The chamber pressure was maintained at maintained at 2mTorr and the PCB temperature was monitored using an RTD. DC I–V  measurements were carried out using two Keithley 237 source-measure units (SMUs) which supplied the drain and gate voltages and measured the drain current. The source and body of the MOSFETs were tied to ground. Data acquisition was automated with a Python script using the PyVISA library for GPIB-based instrument control. The schematic of the cryogenic I–V measurement setup along with the wire bonded chip and the cryo-vacuum system used for the measurement is shown in Figure \ref{fig:hardware_setup}. A noise floor of 100 nA was seen in the cryogenic I-V measurements.

Using the same set of SMUs and data acquisition scripts, room temperature I–V curves for several transistors were also taken to verify the testing configuration and the room temperature models.

\section{Single Transistor I–V Measurements and Characterization}\label{sec:measRes}
\begin{table}[htbp]
\centering
    \begin{tabular}{c c | c c c c}
    \hline
    \multicolumn{2}{c}{\textbf{pMOS Size}} &
    \multicolumn{4}{|c}{\textbf{pMOS SKY130 Model Bins}}\\
    \hline
    Length (L) & Width (W) & $L_{min}$ & $L_{max}$ & $W_{min}$ & $W_{max}$ \\
    \hline 
    0.35 & 0.55 & 0.35 & 0.5 & 0.42 & 0.55 \\
    0.35 & 1.6 & 0.35 & 0.5 & 1 & 3 \\
    0.35 & 5 & 0.35 & 0.5 & 3 & 5 \\
    0.5 & 0.55 & 0.35 & 0.5 & 0.55 & 1 \\
    0.5 & 0.64 & 0.5 & 1 & 0.55 & 1 \\
    2 & 5 & 2 & 4 & 3 & 5 \\
    4 & 7 & 4 & 8 & 5 & 7 \\
    8 & 0.84 & 4 & 8 & 0.55 & 1 \\
    8 & 1.6 & 4 & 8 & 1 & 3 \\
    8 & 5 & 4 & 8 & 3 & 5 \\
    \hline
    \end{tabular}
    \caption{PMOS geometries tested at 77\,K and the respective model bins used to generate the 77\,K model. Lengths and widths are in micrometers.}
    \label{tab:pmos}
\end{table}
\begin{table}[htbp]
    \centering
    \begin{tabular}{c c | c c c c}
    \hline
    \multicolumn{2}{c}{\textbf{nMOS Size}} &
    \multicolumn{4}{|c}{\textbf{nMOS SKY130 Model Bins}}\\
    \hline
    Length (L) & Width (W) & $L_{min}$ & $L_{max}$ & $W_{min}$ & $W_{max}$\\
    \hline 
    0.15 & 1.6 & 0.15 & 0.18 & 1 & 1.65 \\
    0.19 & 7 & 0.18 & 0.25 & 5.05 & 7 \\
    0.25 & 1.6 & 0.18 & 0.25 & 1 & 1.65 \\
    1 & 1.6 & 1 & 2 & 1 & 1.65 \\
    1 & 3 & 0.5 & 1 & 1.65 & 3 \\
    8 & 1.6 & 4 & 8 & 1 & 1.65 \\
    20 & 0.64 & 8 & 100 & 0.55 & 0.64 \\
    100 & 100 & 8 & 100 & 7 & 100 \\
    \hline
    \end{tabular}
    \caption{NMOS geometries tested at 77\,K and the respective model bins used to generate the 77\,K model. Lengths and widths are in micrometers.}
    \label{tab:nmos}
\end{table}
At 77\,K, output and transfer characteristics were measured for twenty-two transistors, eight nMOS and fourteen pMOS, over a range of geometries. In the SKY130 BSIM4 model set, parameter values are binned by channel length and width, such that devices with different geometries are described by different model bins. The measured devices, along with their channel dimensions and associated bin assignments, are listed in Tables~\ref{tab:pmos} and \ref{tab:nmos} for pMOS and nMOS transistors, respectively. A device with geometry $(L,W)$ is assigned to the bin
\[
L \in (L_{\min},L_{\max}] \quad \text{and} \quad W \in (W_{\min},W_{\max}] .
\] However, if that model bin has already been assigned to a different transistor, the neighboring bin is chosen to maximize the number of unique cryogenic models developed.

\begin{figure}[htbp]
\centering

\begin{subfigure}{0.45\textwidth}
    \centering
    \includegraphics[width=\textwidth]{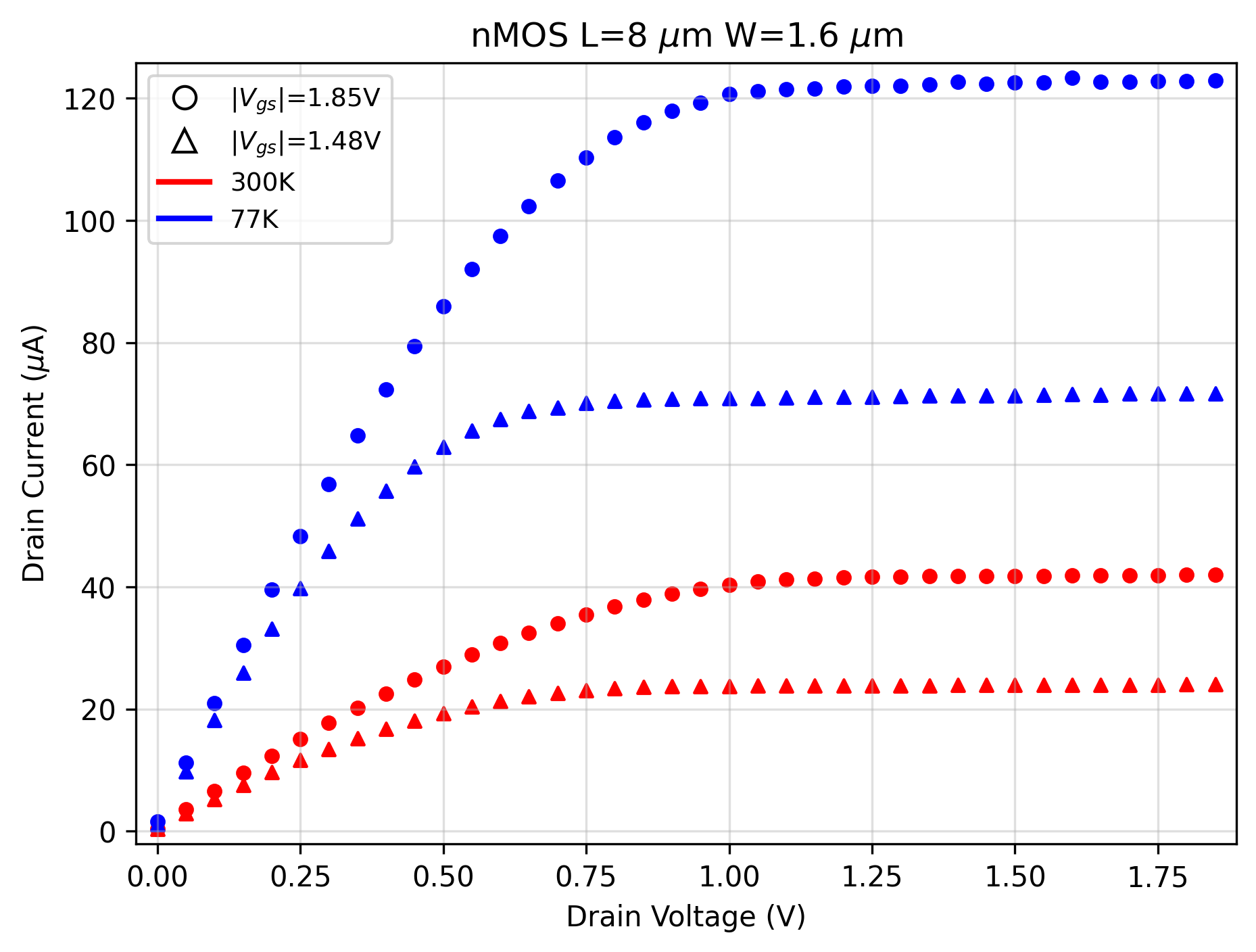}
    \caption{}
    \label{fig:nmos_idvd}
\end{subfigure}
\hfill
\begin{subfigure}{0.495\textwidth}
    \centering
    \includegraphics[width=\textwidth]{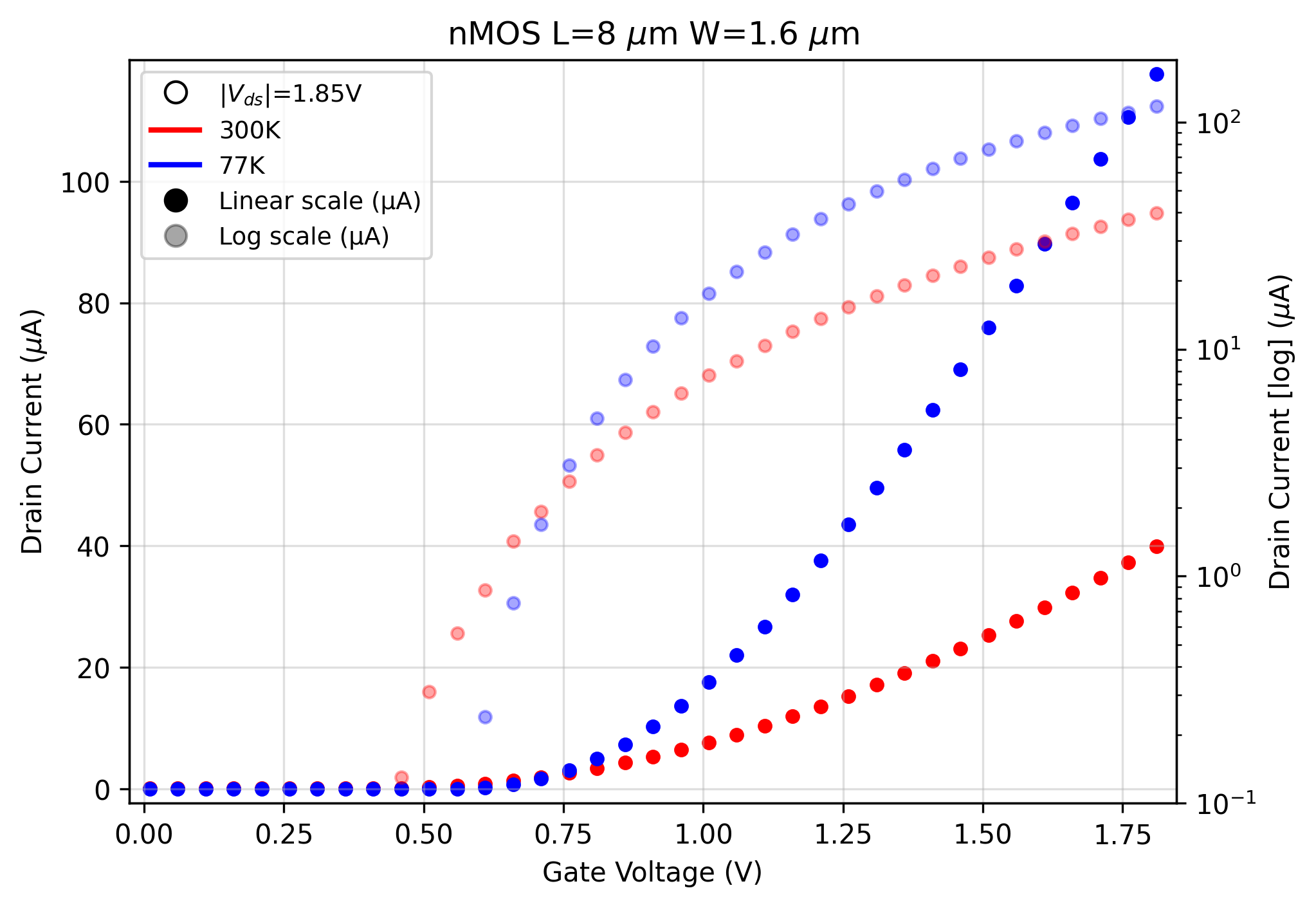}
    \caption{}
    \label{fig:nmos_idvg}
\end{subfigure}

\vspace{0.4cm}

\begin{subfigure}{0.45\textwidth}
    \centering
    \includegraphics[width=\textwidth]{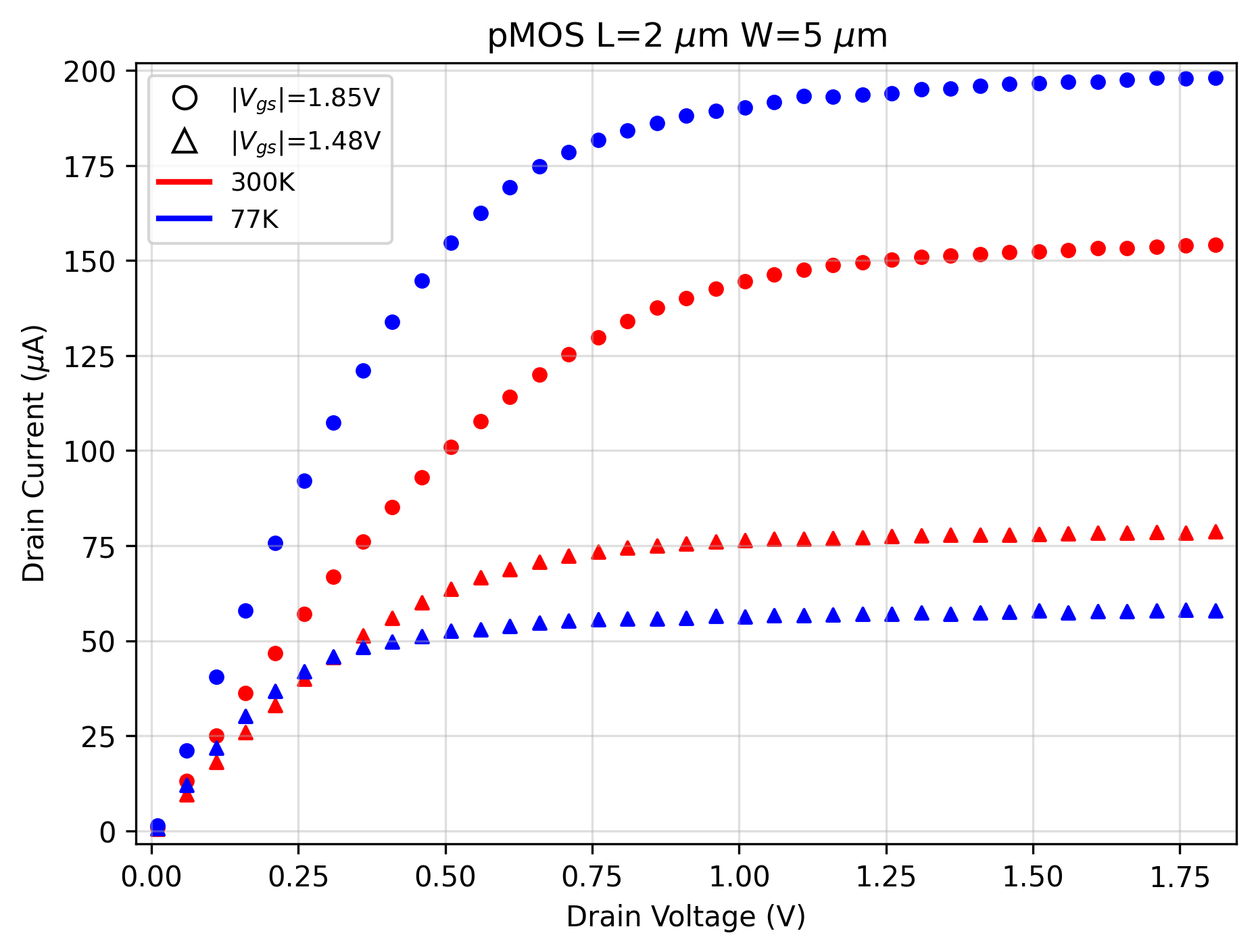}
    \caption{}
    \label{fig:pmos_idvd}
\end{subfigure}
\hfill
\begin{subfigure}{0.492\textwidth}
    \centering
    \includegraphics[width=\textwidth]{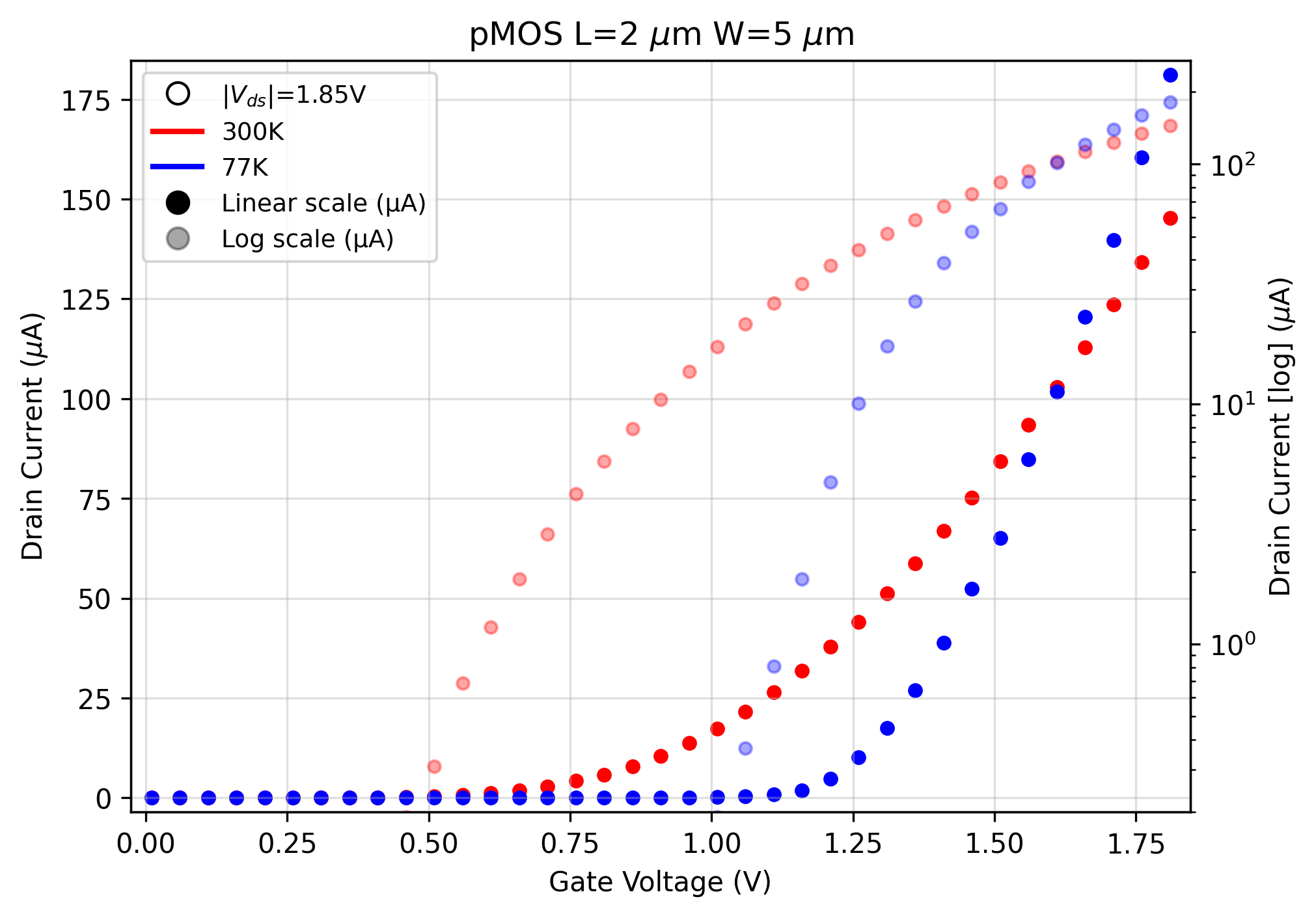}
    \caption{}
    \label{fig:pmos_idvg}
\end{subfigure}

\caption{I–V characteristics measured at 77\,K and 300\,K: (a) Drain current versus Drain voltage ($I_D$--$V_{DS}$) for two select $V_{GS}$ biases and (b) Drain current versus Gate voltage ($I_D$--$V_{GS}$) (linear and log scale) for one $V_{DS}$ bias for nMOS of length 8 $\mu$m and width 1 $\mu$m. (c) $I_D$--$V_{DS}$ for two select $V_{GS}$ biases and (d) $I_D$--$V_{GS}$ (linear and log scale) for one $V_{DS}$ bias for pMOS of length 2 $\mu$m and width 4 $\mu$m. The absolute value of voltage and current values are displayed. The source and body bias are fixed at ground for all measurements.}
\label{fig:RT_vs_cold}
\end{figure}

A selected set of measured output and transfer characteristic curves for representative MOSFETs comparing their 300\,K and 77\,K data are shown in Figure \ref{fig:RT_vs_cold}. Both transistors show an increase in maximum drain current at 77\,K from the 300\,K value at $V_{GS}=V_{DS}=V_{max}=1.85$V for this process, consistent with previous results \cite{cryo28QNTMBeckers2017}. The steepness of the subthreshold slope at 77\,K is clearly visible in the logarithmic transfer curves of Figure \ref{fig:nmos_idvg} and \ref{fig:pmos_idvg}, consistent with previous results \cite{cryo28QNTMBeckers2017,sub4KKamgar1982,28BulkBeckers2018,nanoSCBalestra2017}.

The complete set of 300\,K and 77\,K data is provided in a publicly available Github repository~\cite{github}. The following section provides the cryogenic modeling methodology and results.

\section{Cryogenic Modeling} \label{sec:cryomodeling}
\subsection{Methodology}

The SKY130 room temperature PDK used in this work was downloaded from Volare through Open\_PDKs for the most recent model version available at the time of building the 77\,K, cryogenic model (version number listed in SKY130 77\,K model Github in the README file) ~\cite{Volare, Open_PDKs}. The foundry-verified, room temperature PDK is the foundation on which the 77\,K model is built, assuming all transistors are typical-typical process corner.  In order to construct an isothermal, cryogenic PDK at 77\,K, we first change the model parameter $T_{NOM}$ from its default, room temperature value to 77\,K ($-196.15\degree$C). This parameter identifies the temperature at which the PDK was originally validated, which is 30$\degree$C in the case of SKY130. Then, a physics-based parameter extraction strategy is devised, guided by prior BSIM4 cryogenic modeling efforts (\cite{akturk2023cryogenicsky1304k,65nmcryoGatti2024, deepcryomodelIncandela2018,cryoBSIMLu2021}), but modified for our purposes. Following the methodology of previous work, only DC characteristics are used, since previous C--V studies in advanced CMOS nodes suggest that the dominant
low-temperature effect on intrinsic gate-capacitance ($C_{gg}$) characteristics
is often a shift in the bias axis associated with $V_{TH}(T)$, with
comparatively smaller changes in the overall shape of $C_{gg}(V_G)$. \cite{28BulkBeckers2018}. Parameter extraction for cryogenic modeling at 77\,K is therefore based on the temperature-sensitive BSIM4 parameters identified in Section \ref{sec:cryoCMOS}. 

The Synopsys tool, Mystic\texttrademark{} (Version 2023.12), is utilized to perform parameter extraction \cite{mystic}. Mystic\texttrademark{} is a Python-based optimizer compatible with HSPICE\textregistered{} \cite{HSPICE}. The Mystic\texttrademark{} tool allows for target data, simulated model data, parameter sets, and select I–V curve(s) to target for extraction as inputs for an extraction stage and seeks to find optimal solutions by adjusting specified model parameters to achieve best fit of the target data and model I–V curves.  

We employ a unique approach to parameter extraction, whereby instead of extracting the literal BSIM4 parameters, we extract nominal parameters (defaulted to one) that are manually inserted into the PDK. These are multiplicative factors to the original BSIM4 parameter. This strategy preserves the underlying PDK dynamics of the original BSIM4 parameters, such as mismatch and process corner variations. The nominal parameter scales the original BSIM4 parameter to the new 77\,K values. This method is advantageous in cryogenic modeling, since it's difficult to obtain cryogenic data for all fabrication variations.

\subsection{Model and Design Parameters} \label{sec:ModelAndDesignParameters}
The extracted parameter set and fitting order were chosen to target the cryogenic transistor effects described in Section \ref{sec:cryoCMOS}. Here we summarize only the extraction strategy and the I–V regions used for fitting. Typical extraction methodologies employ a multi-stage optimization routine which extracts small groups of parameters against select regions of the I–V curves. The staged procedure reduces the number of parameters adjusted simultaneously, but it does not
guarantee unique identification of their physical contributions.
The fitted values are interpreted as an effective compact-model parameterization constrained by the measured current--voltage characteristics.

The SKY130 models are binned by length and width with BSIM4 parameters varying across length and width bins. The chosen parameters are extracted at 77\,K per bin against I–V data from devices tested at 77\,K which fall into the particular bins. 

\begin{table}[htbp]
\centering
\small
\renewcommand{\arraystretch}{1.3}
\begin{tabularx}{\textwidth}{l| X| X| X}
\toprule
\textbf{Parameter} & \textbf{Description} & \textbf{Expected Trend From 300\,K to 77\,K} & \textbf{Extraction Location} \\
\midrule
\verb=VTH0= & Threshold voltage & Increase (see Section~\ref{vth_section}) & $I_D$--$V_{GS}$ ($|V_{DS}| = 0.01\,\mathrm{V}$) \\

\verb=U0= & Low-field mobility & Increase (see Section~\ref{mobility_section}) & $I_D$--$V_{GS}$ ($|V_{DS}| = 0.01\,\mathrm{V}$) for nMOS, $I_D$--$V_{GS}$ ($|V_{DS}| = 1.85\,\mathrm{V}$) for pMOS (see Section~\ref{sec:ModelAndDesignParameters}) \\

\verb=RDSW= & LDD source/drain resistance & Increase/Decrease (see Section~\ref{Rsd_section}) & $I_D$--$V_{GS}$ ($|V_{DS}| = 0.01\,\mathrm{V}$) \\

\verb=NFACTOR= & Subthreshold swing factor & Increase (see Section~\ref{SS_section}) & $I_D$--$V_{GS}$ ($|V_{DS}| = 0.01\,\mathrm{V}$) \\

\verb=VSAT= & Saturation velocity & Increase (see Section~\ref{Ion_section}) & $I_D$--$V_{GS}$ ($|V_{DS}| = 1.85\,\mathrm{V}$), $I_D$--$V_{DS}$ \\

\verb=ETA0= & DIBL coefficient & Increase (if needed) (see Section~\ref{dibl_section}) & $I_D$--$V_{GS}$ ($|V_{DS}| = 1.85\,\mathrm{V}$) \\

\verb=DELTA= & Linear to saturation smoothing & Increase (if needed) (see Section~\ref{Ion_section}) & $I_D$--$V_{DS}$ \\

\bottomrule
\end{tabularx}
\caption{Extracted BSIM4 parameters at 77\,K with descriptions, expected trends relative to 300\,K, and extraction data regions.}
\label{fig:parameters}
\end{table}

Table~\ref{fig:parameters} lists the extracted parameters and the I–V curve(s) used for extraction. Stringent parameter boundaries are implemented to aid in optimization time and accuracy. For nMOS devices, the first four parameters (\verb=VTH0=, \verb=U0=, \verb=RDSW=, and \verb=NFACTOR=) are extracted against the lowest drain-to-source voltage ($V_{DS}$) biased transfer characteristic curve ($|V_{DS}|=0.01$V, in this case). Next, \verb=VSAT= was extracted against all $I_D$--$V_{DS}$ curves and the highest drain voltage biased $I_D$--$V_{GS}$ curve. The drain-induced barrier lowering (DIBL) parameter \verb=ETA0= was introduced for select, short length devices that exhibited DIBL behavior. \verb=ETA0= is extracted against the $I_D$--$V_{GS}$ curve with the maximum $V_{DS}$ (in this case, $V_{DS}=1.85$V).
A similar approach was used for the parameter, \verb=DELTA=, which is introduced to ensure a smooth transition from the linear to saturation operating region when the fit in that data region is visually poor. \verb=DELTA= is extracted against all $I_D$--$V_{DS}$ curves. A pMOS-specific calibration procedure was adopted because applying
the initial four-parameter low-drain-bias optimization to the pMOS devices produced good agreement at $|V_{DS}|=0.01~\mathrm{V}$ but
under predicted the higher-drain-bias transfer and output currents. For the retained pMOS calibration, \texttt{VTH0}, \texttt{RDSW}, and \texttt{NFACTOR} were determined from the low-drain-bias transfer characteristic and held fixed while \texttt{U0} was subsequently
calibrated using the $I_D$--$V_{GS}$ characteristic at
$|V_{DS}|=1.85~\mathrm{V}$. This choice improved agreement with the higher-drain-bias characteristics at the expense of an over prediction
of the current at $|V_{DS}|=0.01~\mathrm{V}$.

The low-drain-bias data constrain the turn-on behavior, subthreshold
slope, and low-field conduction, but do not independently determine
all of the underlying physical parameters. In particular,
\texttt{U0} and \texttt{RDSW} both influence terminal current. Their final fitted values therefore represent a calibration trade-off rather than independent measurements of physical mobility and source/drain resistance.

\begin{figure}[htbp]
\centering
\includegraphics[width=0.99\linewidth]{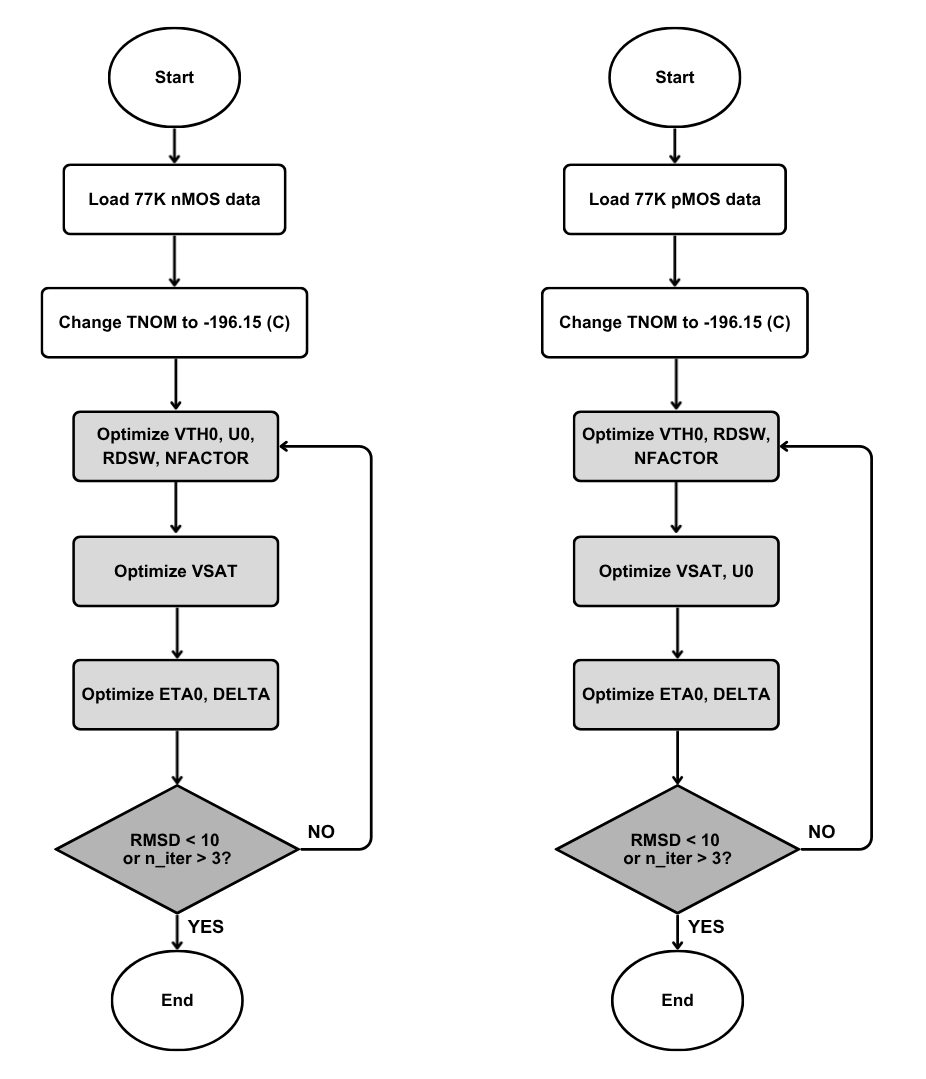}
\caption{Parameter optimization flow chart for nMOS and pMOS isothermal cryogenic model generation.}
\label{fig:flow_chart}
\end{figure}

The above extraction methodology was repeated for every transistor which had a unique length/width bin, following the flow of Figure \ref{fig:flow_chart}. The optimization loop repeats and a root mean square deviation (RMSD) is calculated after every iteration. The loop continues until it reaches a specified number of iterations (in this case, three) or the RMSD of the chosen dataset calculated falls below 10 $\mu$A, which is less than 10\% error in the high current region across all devices. 

\subsection{Modeling Results}
Due to the length and width binning of the SKY130 PDK, several of the twenty-two transistors tested at 77\,K have geometries which fall in the same geometry bin. This yields eighteen distinct 77\,K models. Table \ref{tab:bsim4_params} provides the relevant BSIM4 parameters for these models, including the extracted 77\,K values and the 300\,K (unchanged) values for comparison.

\begin{longtable}{cc}
\caption{Tables of 300\,K room temperature and 77\,K extracted BSIM4 model parameters for SKY130 for various types and length and width bins. \textbf{Note}: \texttt{VTH0} and \texttt{NFACTOR} parameters are listed in a functional form in the model, (refer to the BSIM4 manual \cite{bsim4manual} for their explicit forms). Symbols * and $\dagger$ are placeholders for these equations. The 77\,K value is the functional form times the extracted nominal parameter.}

\label{tab:bsim4_params} \\
\newpage
\multicolumn{2}{c}{\large\textbf{nMOS}}\\[0.5em] \\
    \begin{minipage}{0.48\linewidth}
    \centering
    \begin{tabular}{|c|c c|}
    \multicolumn{3}{c}{\textbf{Lmin = 0.15, Lmax = 0.18}} \\
    \multicolumn{3}{c}{\textbf{Wmin = 1, Wmax = 1.65}}\\
    \hline
    \multicolumn{1}{|c|}{Parameter} & \multicolumn{1}{c}{300\,K} & \multicolumn{1}{c|}{77\,K} \\
    \hline
        \verb=VTH0= & (*) & (*) $\times$ 1.29 \\
        \verb=U0=    & 4.31$\times10^{-2}$ & 7.40$\times10^{-2}$ \\
        \verb=RDSW=  & 104       & 100 \\
        \verb=NFACTOR= & ($\dagger$) & ($\dagger$) $\times$ 32.2  \\
        \verb=VSAT=  & 7.84$\times10^{4}$ & 1.38$\times10^5$ \\
        \verb=ETA0=  & -0.136 & -0.191 \\
        \verb=DELTA= & 0.01          & 0.053 \\
    \hline
    \end{tabular}
    \vspace{1.5em}
    \end{minipage}
    &
    \begin{minipage}{0.48\linewidth}
    \centering
    \begin{tabular}{|c|c c|}
    \multicolumn{3}{c}{\textbf{Lmin = 0.18, Lmax = 0.25}} \\
    \multicolumn{3}{c}{\textbf{Wmin = 1, Wmax = 1.65}}\\
    \hline
    \multicolumn{1}{|c|}{Parameter} & \multicolumn{1}{c}{300\,K} & \multicolumn{1}{c|}{77\,K} \\
    \hline
        \verb=VTH0= & (*) & (*) $\times$ 1.19 \\
        \verb=U0=    & 2.58$\times 10^{-2}$ & 1.06$\times 10^{-1}$ \\
        \verb=RDSW=  & 104 & 158 \\
        \verb=NFACTOR= & ($\dagger$) & ($\dagger$) $\times$ 7.51 \\
        \verb=VSAT=  & 1.24$\times10^5$  & 1.52$\times10^5$ \\
        \verb=ETA0=  & -1.56$\times10^{-1}$ & -2.90$\times10^{-1}$ \\
        \verb=DELTA= & 0.01   & 0.058 \\
    \hline
    \end{tabular}
    \vspace{1.5em}
    \end{minipage}
    
    \\[1.5em]
    \begin{minipage}{0.48\linewidth}
    \centering
    \begin{tabular}{|c|c c|}
    \multicolumn{3}{c}{\textbf{Lmin = 0.18, Lmax = 0.25}} \\
    \multicolumn{3}{c}{\textbf{Wmin = 5.5, Wmax = 7}}\\
    \hline
    \multicolumn{1}{|c|}{Parameter} & \multicolumn{1}{c}{300\,K} & \multicolumn{1}{c|}{77\,K} \\
    \hline
        \verb=VTH0= & (*) & (*) $\times$ 1.12 \\
        \verb=U0=    & 2.11$\times10^{-2}$ & 8.13$\times10^{-2}$ \\
        \verb=RDSW=  & 104       & 143 \\
        \verb=NFACTOR= & ($\dagger$) & ($\dagger$) $\times$ 35  \\
        \verb=VSAT=  & 1.90$\times10^{5}$ & 2.31$\times10^5$ \\
        \verb=ETA0=  & -5.86$\times10^{-2}$ & -2.45$\times10^{-1}$ \\
        \verb=DELTA= & 0.01          & 0.085 \\
    \hline
    \end{tabular}
    \vspace{1.5em}
    \end{minipage}
    &
    \begin{minipage}{0.48\linewidth}
    \centering
    \begin{tabular}{|c|c c|}
    \multicolumn{3}{c}{\textbf{Lmin = 0.5, Lmax = 1}} \\
    \multicolumn{3}{c}{\textbf{Wmin = 1.65, Wmax = 3}}\\
    \hline
    \multicolumn{1}{|c|}{Parameter} & \multicolumn{1}{c}{300\,K} & \multicolumn{1}{c|}{77\,K} \\
    \hline
        \verb=VTH0= & (*) & (*) $\times$ 1.29 \\
        \verb=U0=    & 2.59$\times10^{-2}$ & 2.93$\times10^{-1}$ \\
        \verb=RDSW=  & 104       & 288 \\
        \verb=NFACTOR= & ($\dagger$) & ($\dagger$) $\times$ 8.74  \\
        \verb=VSAT=  & 1.24$\times10^{5}$ & 1.52$\times10^5$ \\
        \verb=ETA0=  & 0.004 & 0.004 \\
        \verb=DELTA= & 0.01          & 0.05 \\
    \hline
    \end{tabular}
    \vspace{1.5em}
    \end{minipage}

\\[1em]
    \begin{minipage}{0.48\linewidth}
    \centering
    \begin{tabular}{|c|c c|}
    \multicolumn{3}{c}{\textbf{Lmin = 1, Lmax = 2}} \\
    \multicolumn{3}{c}{\textbf{Wmin = 1, Wmax = 1.65}}\\
    \hline
    \multicolumn{1}{|c|}{Parameter} & \multicolumn{1}{c}{300\,K} & \multicolumn{1}{c|}{77\,K} \\
    \hline
        \verb=VTH0= & (*) & (*) $\times$ 1.31 \\
        \verb=U0=    & 3.62$\times10^{-2}$ & 2.20$\times10^{-1}$ \\
        \verb=RDSW=  & 104       & 245 \\
        \verb=NFACTOR= & ($\dagger$) & ($\dagger$) $\times$ 6.04  \\
        \verb=VSAT=  & 2.30$\times10^{3}$ & 3.82$\times10^4$ \\
        \verb=ETA0=  & 5.65$\times10^{-4}$ & 5.65$\times10^{-4}$ \\
        \verb=DELTA= & 0.01          & 0.08 \\
    \hline
    \end{tabular}
    \vspace{1.5em}
    \end{minipage}
    
    &
    \begin{minipage}{0.48\linewidth}
    \centering
    \begin{tabular}{|c|c c|}
    \multicolumn{3}{c}{\textbf{Lmin = 4, Lmax = 8}} \\
    \multicolumn{3}{c}{\textbf{Wmin = 1, Wmax = 1.65}}\\
    \hline
    \multicolumn{1}{|c|}{Parameter} & \multicolumn{1}{c}{300\,K} & \multicolumn{1}{c|}{77\,K} \\
    \hline
        \verb=VTH0= & (*) & (*) $\times$ 1.32 \\
        \verb=U0=    & 3.12$\times10^{-2}$ & 2.28$\times10^{-1}$ \\
        \verb=RDSW=  & 104       & 1.04$\times10^{-1}$ \\
        \verb=NFACTOR= & ($\dagger$) & ($\dagger$) $\times$ 2.41  \\
        \verb=VSAT=  & 2.01$\times10^{5}$ & 3.97$\times10^4$ \\
        \verb=ETA0=  & 0.08 & 0.08 \\
        \verb=DELTA= & 0.01          & 0.15 \\
    \hline
    \end{tabular}
    \vspace{1.5em}
\end{minipage}

\\[1em]
    \begin{minipage}{0.48\linewidth}
    \centering
    \begin{tabular}{|c|c c|}
    \multicolumn{3}{c}{\textbf{Lmin = 8, Lmax = 100}} \\
    \multicolumn{3}{c}{\textbf{Wmin = 0.55, Wmax = 0.64}}\\
    \hline
    \multicolumn{1}{|c|}{Parameter} & \multicolumn{1}{c}{300\,K} & \multicolumn{1}{c|}{77\,K} \\
    \hline
        \verb=VTH0= & (*) & (*) $\times$ 1.06 \\
        \verb=U0=    & 3.12$\times10^{-2}$ & 2.90$\times10^{-1}$ \\
        \verb=RDSW=  & 104       & 3.73$\times10^{3}$ \\
        \verb=NFACTOR= & ($\dagger$) & ($\dagger$) $\times$ 23.4  \\
        \verb=VSAT=  & 2.01$\times10^{5}$ & 2.36$\times10^4$ \\
        \verb=ETA0=  & 0.08 & 0.08 \\
        \verb=DELTA= & 0.01          & 0.150 \\
    \hline
    \end{tabular}
    \vspace{1.5em}
    \end{minipage}
    
    &
    \begin{minipage}{0.48\linewidth}
    \centering
    \begin{tabular}{|c|c c|}
    \multicolumn{3}{c}{\textbf{Lmin = 8, Lmax = 100}} \\
    \multicolumn{3}{c}{\textbf{Wmin = 7, Wmax = 100}}\\
    \hline
    \multicolumn{1}{|c|}{Parameter} & \multicolumn{1}{c}{300\,K} & \multicolumn{1}{c|}{77\,K} \\
    \hline
        \verb=VTH0= & (*) & (*) $\times$ 1.26 \\
        \verb=U0=    & 3.20$\times10^{-2}$ & 2.90$\times10^{-1}$ \\
        \verb=RDSW=  & 104       & 1.50$\times10^{4}$ \\
        \verb=NFACTOR= & ($\dagger$) & ($\dagger$) $\times$ 4.31  \\
        \verb=VSAT=  & 1.61$\times10^{5}$ & 6.54$\times10^3$ \\
        \verb=ETA0=  & 0.08 & 0.08 \\
        \verb=DELTA= & 0.01          & 0.1 \\
    \hline
    \end{tabular}
    \vspace{1.5em}
\end{minipage}

\\[2em]
\multicolumn{2}{c}{\large\textbf{pMOS}}\\[0.5em] \\
    \begin{minipage}{0.48\linewidth}
    \centering
    \begin{tabular}{|c|c c|}
    \multicolumn{3}{c}{\textbf{Lmin = 0.35, Lmax = 0.5}} \\
    \multicolumn{3}{c}{\textbf{Wmin = 0.42, Wmax = 0.55}}\\
    \hline
    \multicolumn{1}{|c|}{Parameter} & \multicolumn{1}{c}{300\,K} & \multicolumn{1}{c|}{77\,K} \\
    \hline
        \verb=VTH0= & (*) & (*) $\times$ 2.84 \\
        \verb=U0=    & 2.84$\times10^{-3}$ & 3.10$\times10^{-3}$ \\
        \verb=RDSW=  & 485       & 6.43 \\
        \verb=NFACTOR= & ($\dagger$) & ($\dagger$) $\times$ 28.8  \\
        \verb=VSAT=  & -4.91$\times10^{5}$ & -2.59$\times10^4$ \\
        \verb=ETA0=  & 0.2 & 0.2 \\
        \verb=DELTA= & 0.128 & 0.128 \\
    \hline
    \end{tabular}
    \vspace{1.5em}
    \end{minipage}
    &
    \begin{minipage}{0.48\linewidth}
    \centering
    \begin{tabular}{|c|c c|}
    \multicolumn{3}{c}{\textbf{Lmin = 0.35, Lmax = 0.5}} \\
    \multicolumn{3}{c}{\textbf{Wmin = 0.55, Wmax = 1}}\\
    \hline
    \multicolumn{1}{|c|}{Parameter} & \multicolumn{1}{c}{300\,K} & \multicolumn{1}{c|}{77\,K} \\
    \hline
        \verb=VTH0= & (*) & (*) $\times$ 4.85 \\
        \verb=U0=    & 9.34$\times10^{-4}$ & 9.33$\times10^{-3}$ \\
        \verb=RDSW=  & 485       & 2.14$\times10^{3}$ \\
        \verb=NFACTOR= & ($\dagger$) & ($\dagger$) $\times$ 16.7  \\
        \verb=VSAT=  & 1.73$\times10^{5}$ & 7.11$\times10^8$ \\
        \verb=ETA0=  & 0.2 & 0.2 \\
        \verb=DELTA= & 0.151 & 0.151 \\
    \hline
    \end{tabular}
    \vspace{1.5em}
    \end{minipage}

\\[1em]
    \begin{minipage}{0.48\linewidth}
    \centering
    \begin{tabular}{|c|c c|}
    \multicolumn{3}{c}{\textbf{Lmin = 0.35, Lmax = 0.5}} \\
    \multicolumn{3}{c}{\textbf{Wmin = 1, Wmax = 3}}\\
    \hline
    \multicolumn{1}{|c|}{Parameter} & \multicolumn{1}{c}{300\,K} & \multicolumn{1}{c|}{77\,K} \\
    \hline
        \verb=VTH0= & (*) & (*) $\times$ 2.08 \\
        \verb=U0=    & 3.49$\times10^{-3}$ & 1.19$\times10^{-2}$ \\
        \verb=RDSW=  & 485       & 5.40$\times10^{-3}$ \\
        \verb=NFACTOR= & ($\dagger$) & ($\dagger$) $\times$ 13.5  \\
        \verb=VSAT=  & 1.03$\times10^{5}$ & 1.03$\times10^9$ \\
        \verb=ETA0=  & 0.2 & 0.2 \\
        \verb=DELTA= & -1.54$\times10^{-2}$ & -1.54$\times10^{-2}$ \\
    \hline
    \end{tabular}
    \vspace{1.5em}
    \end{minipage}
    &
    \begin{minipage}{0.48\linewidth}
    \centering
    \begin{tabular}{|c|c c|}
    \multicolumn{3}{c}{\textbf{Lmin = 0.35, Lmax = 0.5}} \\
    \multicolumn{3}{c}{\textbf{Wmin = 3, Wmax = 5}}\\
    \hline
    \multicolumn{1}{|c|}{Parameter} & \multicolumn{1}{c}{300\,K} & \multicolumn{1}{c|}{77\,K} \\
    \hline
        \verb=VTH0= & (*) & (*) $\times$ 3.01 \\
        \verb=U0=    & 2.52$\times10^{-3}$ & 2.99$\times10^{-2}$ \\
        \verb=RDSW=  & 485       & 2.99$\times10^{3}$ \\
        \verb=NFACTOR= & ($\dagger$) & ($\dagger$) $\times$ 50  \\
        \verb=VSAT=  & 6.58$\times10^{5}$ & 3.09$\times10^9$ \\
        \verb=ETA0=  & 0.2 & 0.2 \\
        \verb=DELTA= & 5.92$\times10^{-2}$ & 5.92$\times10^{-2}$ \\
    \hline
    \end{tabular}
    \vspace{1.5em}
    \end{minipage}

\\[1em]
    \begin{minipage}{0.48\linewidth}
    \centering
    \begin{tabular}{|c|c c|}
    \multicolumn{3}{c}{\textbf{Lmin = 0.5, Lmax = 1}} \\
    \multicolumn{3}{c}{\textbf{Wmin = 0.55, Wmax = 1}}\\
    \hline
    \multicolumn{1}{|c|}{Parameter} & \multicolumn{1}{c}{300\,K} & \multicolumn{1}{c|}{77\,K} \\
    \hline
        \verb=VTH0= & (*) & (*) $\times$ 2.70 \\
        \verb=U0=    & 2.46$\times10^{-3}$ & 6.16$\times10^{-2}$ \\
        \verb=RDSW=  & 485       & 7.27$\times10^3$ \\
        \verb=NFACTOR= & ($\dagger$) & ($\dagger$) $\times$ 27.8  \\
        \verb=VSAT=  & 1.86$\times10^{5}$ & 1.49$\times10^9$ \\
        \verb=ETA0=  & 0.2 & 0.2 \\
        \verb=DELTA= & 3.61$\times10^{-2}$ & 3.61$\times10^{-2}$ \\
    \hline
    \end{tabular}
    \vspace{1.5em}
    \end{minipage}
    &
    \begin{minipage}{0.48\linewidth}
    \centering
    \begin{tabular}{|c|c c|}
    \multicolumn{3}{c}{\textbf{Lmin = 2, Lmax = 4}} \\
    \multicolumn{3}{c}{\textbf{Wmin = 3, Wmax = 5}}\\
    \hline
    \multicolumn{1}{|c|}{Parameter} & \multicolumn{1}{c}{300\,K} & \multicolumn{1}{c|}{77\,K} \\
    \hline
        \verb=VTH0= & (*) & (*) $\times$ 2.55 \\
        \verb=U0=    & 2.38$\times10^{-3}$ & 0.116 \\
        \verb=RDSW=  & 485       & 1.19$\times10^4$ \\
        \verb=NFACTOR= & ($\dagger$) & ($\dagger$) $\times$ 23  \\
        \verb=VSAT=  & 1.24$\times10^{5}$ & 6.70$\times10^6$ \\
        \verb=ETA0=  & 0.2 & 0.2 \\
        \verb=DELTA= & 1.27$\times10^{-2}$ & 1.27$\times10^{-2}$ \\
    \hline
    \end{tabular}
    \vspace{1.5em}
    \end{minipage}

\\[1em]
    \begin{minipage}{0.48\linewidth}
    \centering
    \begin{tabular}{|c|c c|}
    \multicolumn{3}{c}{\textbf{Lmin = 4, Lmax = 8}} \\
    \multicolumn{3}{c}{\textbf{Wmin = 0.55, Wmax = 1}}\\
    \hline
    \multicolumn{1}{|c|}{Parameter} & \multicolumn{1}{c}{300\,K} & \multicolumn{1}{c|}{77\,K} \\
    \hline
        \verb=VTH0= & (*) & (*) $\times$ 2.63 \\
        \verb=U0=    & 2.75$\times10^{-3}$ & 3.19$\times10^{-2}$ \\
        \verb=RDSW=  & 485       & 2.87$\times10^4$ \\
        \verb=NFACTOR= & ($\dagger$) & ($\dagger$) $\times$ 15.5  \\
        \verb=VSAT=  & 1.24$\times10^5$ & 6.11$\times10^8$ \\
        \verb=ETA0=  & 0.2 & 0.2 \\
        \verb=DELTA= & -3.87$\times10^{-3}$ & -3.87$\times10^{-3}$ \\
    \hline
    \end{tabular}
    \vspace{1.5em}
    \end{minipage}
    &
    \begin{minipage}{0.48\linewidth}
    \centering
    \begin{tabular}{|c|c c|}
    \multicolumn{3}{c}{\textbf{Lmin = 4, Lmax = 8}} \\
    \multicolumn{3}{c}{\textbf{Wmin = 1, Wmax = 3}}\\
    \hline
    \multicolumn{1}{|c|}{Parameter} & \multicolumn{1}{c}{300\,K} & \multicolumn{1}{c|}{77\,K} \\
    \hline
        \verb=VTH0= & (*) & (*) $\times$ 2.55 \\
        \verb=U0=    & 3.04$\times10^{-3}$ & 2.54$\times10^{-2}$ \\
        \verb=RDSW=  & 485       & 1.21$\times10^4$ \\
        \verb=NFACTOR= & ($\dagger$) & ($\dagger$) $\times$ 11.1  \\
        \verb=VSAT= & 1.24$\times10^{5}$ & 5.45$\times10^8$ \\
        \verb=ETA0=  & 0.2 & 0.2 \\
        \verb=DELTA= & 1.90$\times10^{-2}$ & 1.90$\times10^{-2}$ \\
    \hline
    \end{tabular}
    \vspace{1.5em}
    \end{minipage}

\\[1em]
    \begin{minipage}{0.48\linewidth}
    \centering
    \begin{tabular}{|c|c c|}
    \multicolumn{3}{c}{\textbf{Lmin = 4, Lmax = 8}} \\
    \multicolumn{3}{c}{\textbf{Wmin = 3, Wmax = 5}}\\
    \hline
    \multicolumn{1}{|c|}{Parameter} & \multicolumn{1}{c}{300\,K} & \multicolumn{1}{c|}{77\,K} \\
    \hline
        \verb=VTH0= & (*) & (*) $\times$ 2.63 \\
        \verb=U0=    & 1.90$\times10^{-3}$ & 2.97$\times10^{-2}$ \\
        \verb=RDSW=  & 485       & 1.57$\times10^{4}$ \\
        \verb=NFACTOR= & ($\dagger$) & ($\dagger$) $\times$ 10.6  \\
        \verb=VSAT=  & 1.24$\times10^{5}$ & 7.43$\times10^8$ \\
        \verb=ETA0=  & 0.2 & 0.2 \\
        \verb=DELTA= & 1.18$\times10^{-2}$          & 1.18$\times10^{-2}$ \\
    \hline
    \end{tabular}
    \vspace{1.5em}
    \end{minipage}
    &
    \begin{minipage}{0.48\linewidth}
    \centering
    \begin{tabular}{|c|c c|}
    \multicolumn{3}{c}{\textbf{Lmin = 4, Lmax = 8}} \\
    \multicolumn{3}{c}{\textbf{Wmin = 5, Wmax = 7}}\\
    \hline
    \multicolumn{1}{|c|}{Parameter} & \multicolumn{1}{c}{300\,K} & \multicolumn{1}{c|}{77\,K} \\
    \hline
        \verb=VTH0= & (*) & (*) $\times$ 2.39 \\
        \verb=U0=    & 2.37$\times10^{-3}$ & 3.73$\times10^{-2}$ \\
        \verb=RDSW=  & 485       & 1.03$\times10^4$ \\
        \verb=NFACTOR= & ($\dagger$) & ($\dagger$) $\times$ 15.3  \\
        \verb=VSAT=  & 1.24$\times10^{5}$ & 7.39$\times10^8$ \\
        \verb=ETA0=  & 0.2 & 0.2 \\
        \verb=DELTA= & -1.19$\times10^{-2}$ & -1.19$\times10^{-2}$ \\
    \hline
    \end{tabular}
    \vspace{1.5em}
\end{minipage}

\\

\end{longtable}

\begin{figure}
\centering

\begin{subfigure}{0.45\textwidth}
    \includegraphics[width=\linewidth]{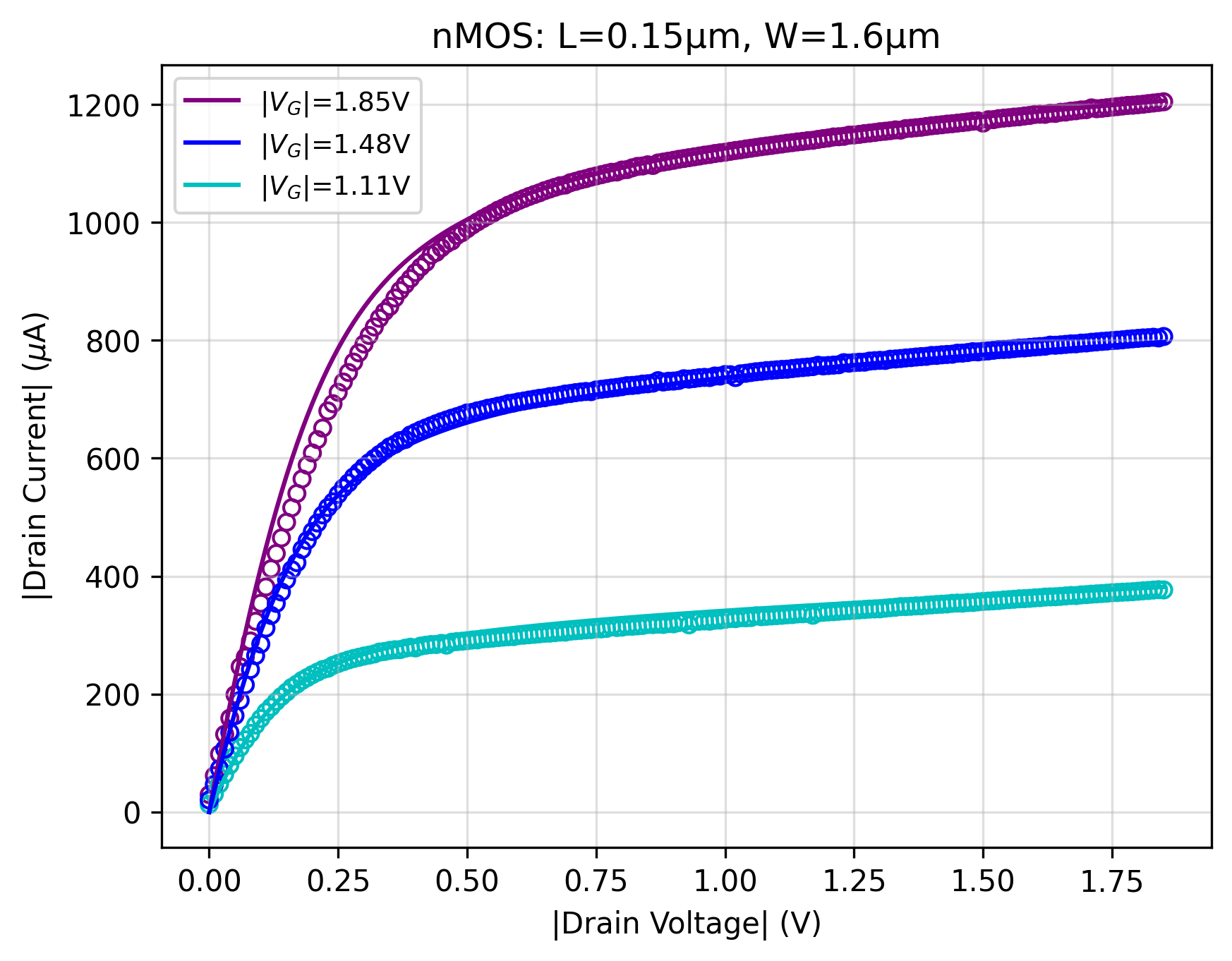}
    \caption{}
\end{subfigure}
\hfill
\begin{subfigure}{0.50\textwidth}
    \includegraphics[width=\linewidth]{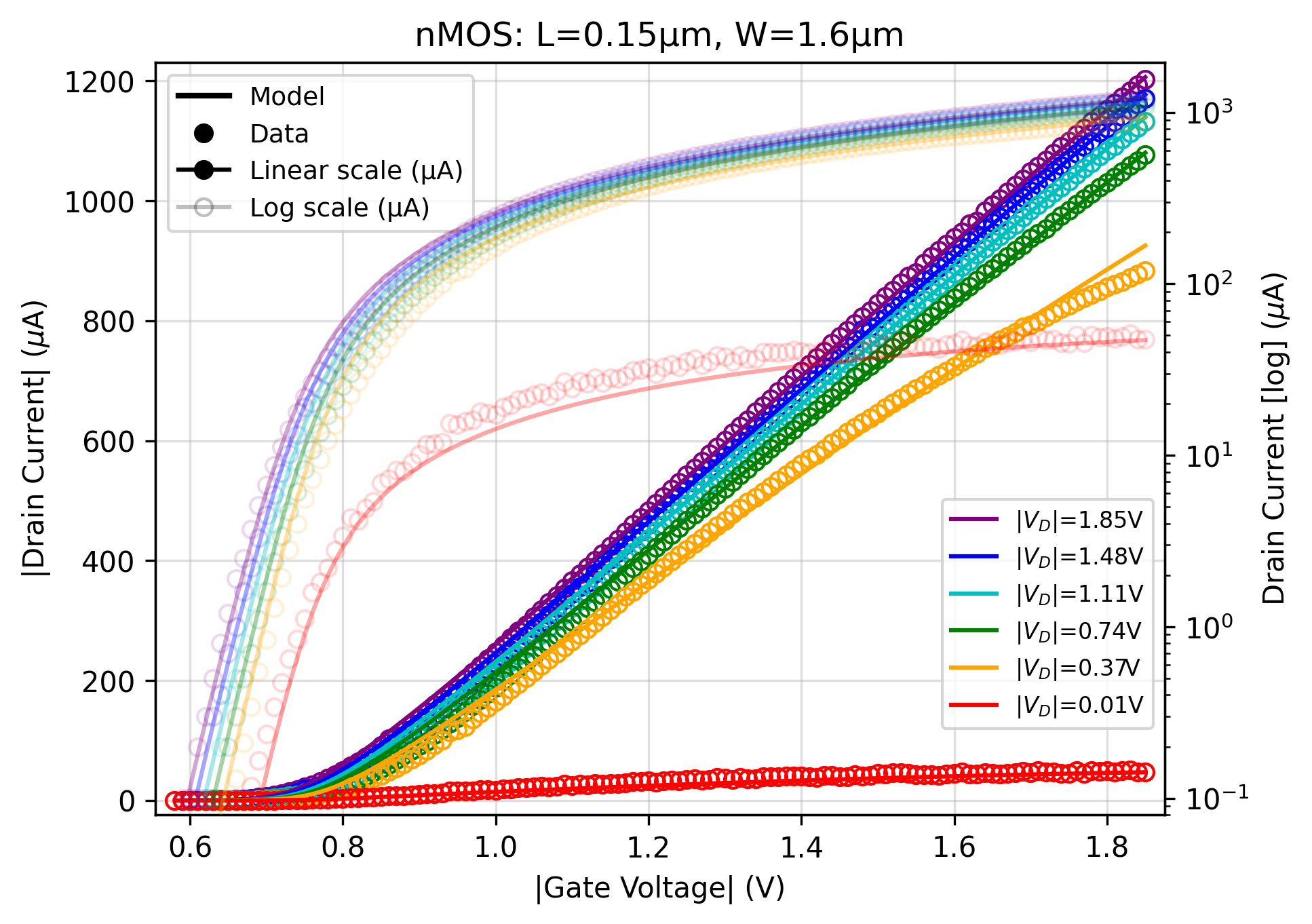}
    \caption{}
\end{subfigure}

\vspace{0.5cm}

\begin{subfigure}{0.45\textwidth}
    \includegraphics[width=\linewidth]{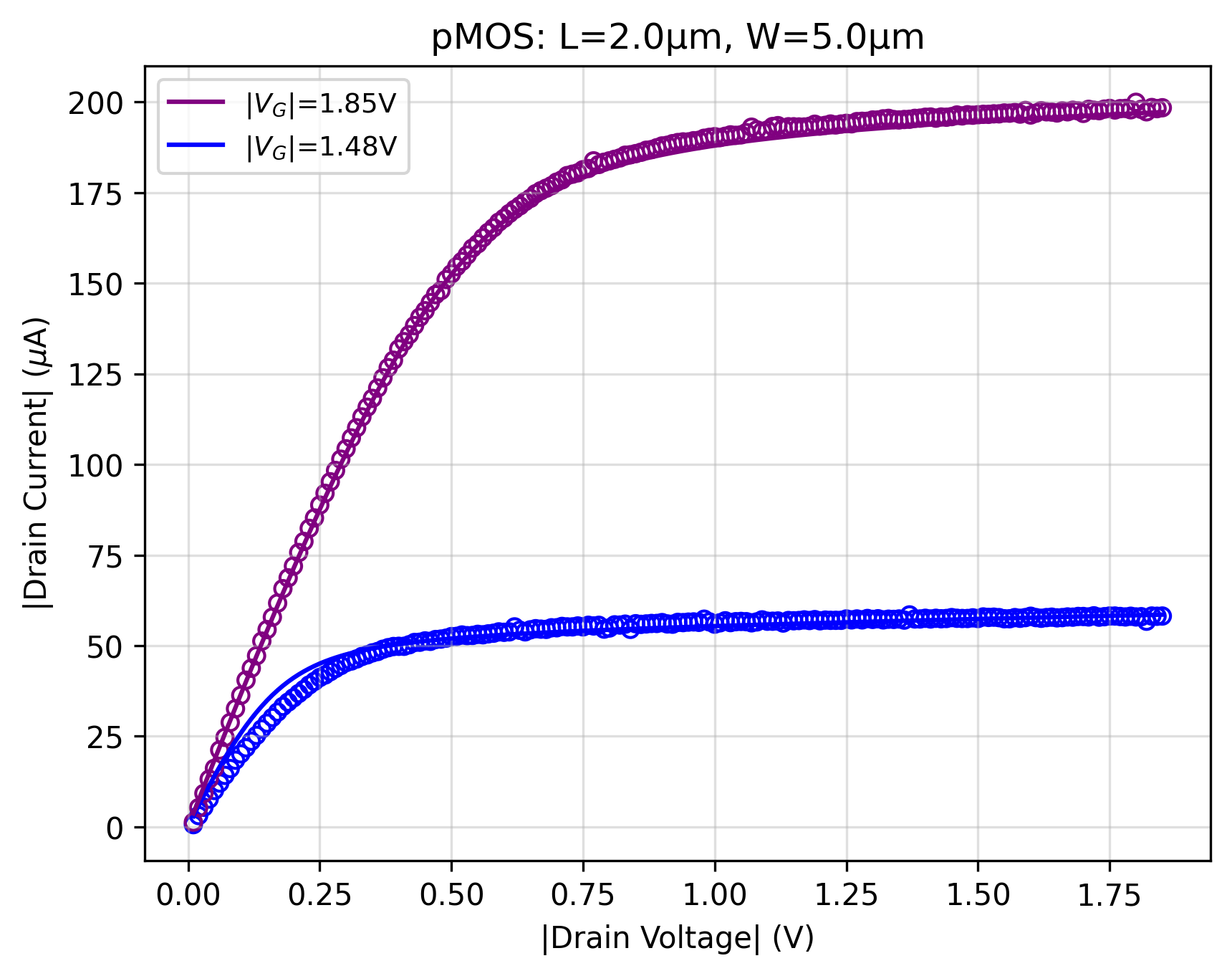} 
    \caption{}
\end{subfigure}
\hfill
\begin{subfigure}{0.50\textwidth}
    \includegraphics[width=\linewidth]{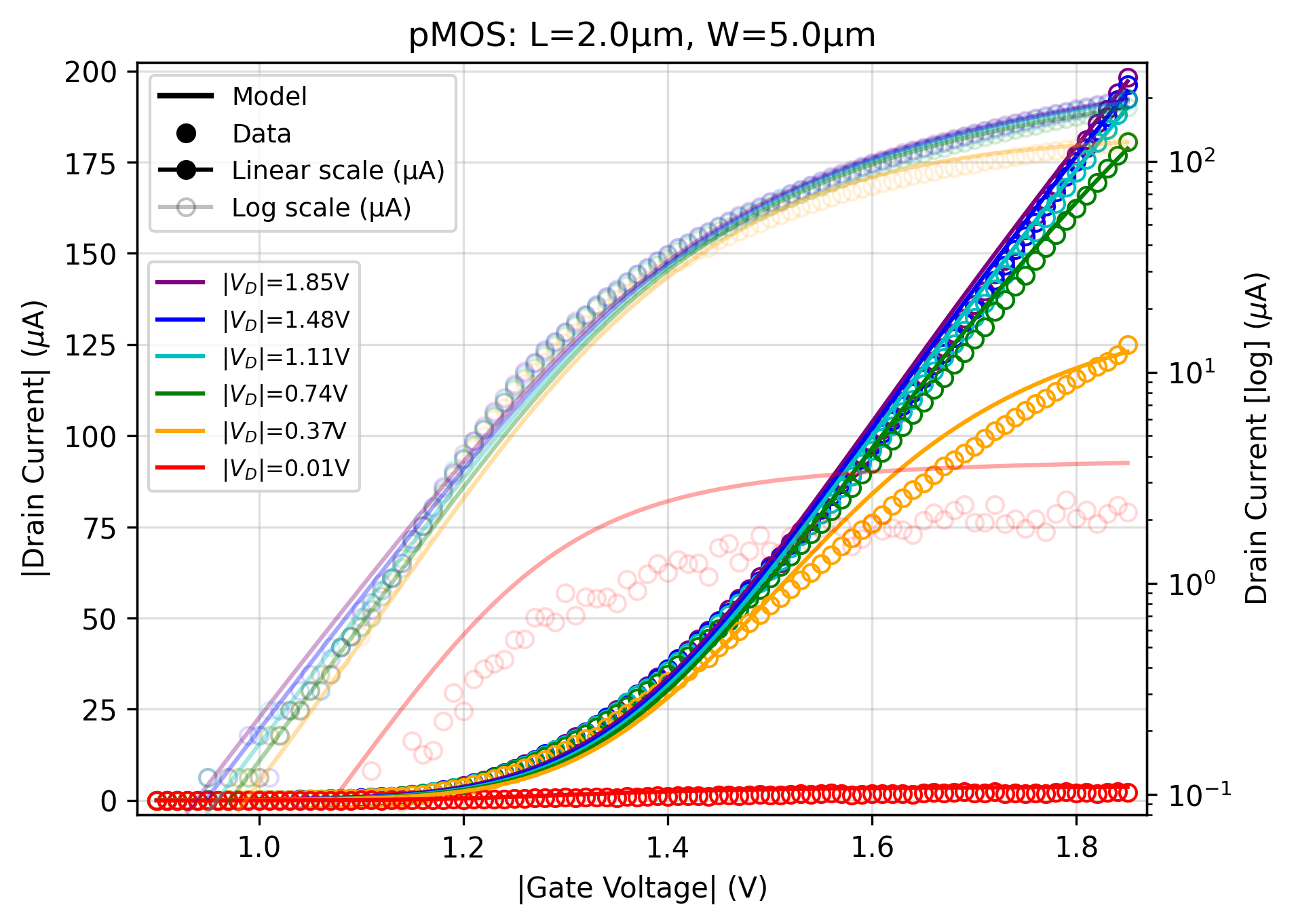} 
    \caption{}
\end{subfigure}

\vspace{0.5cm}

\begin{subfigure}{0.22\textwidth}
    \includegraphics[width=\linewidth]{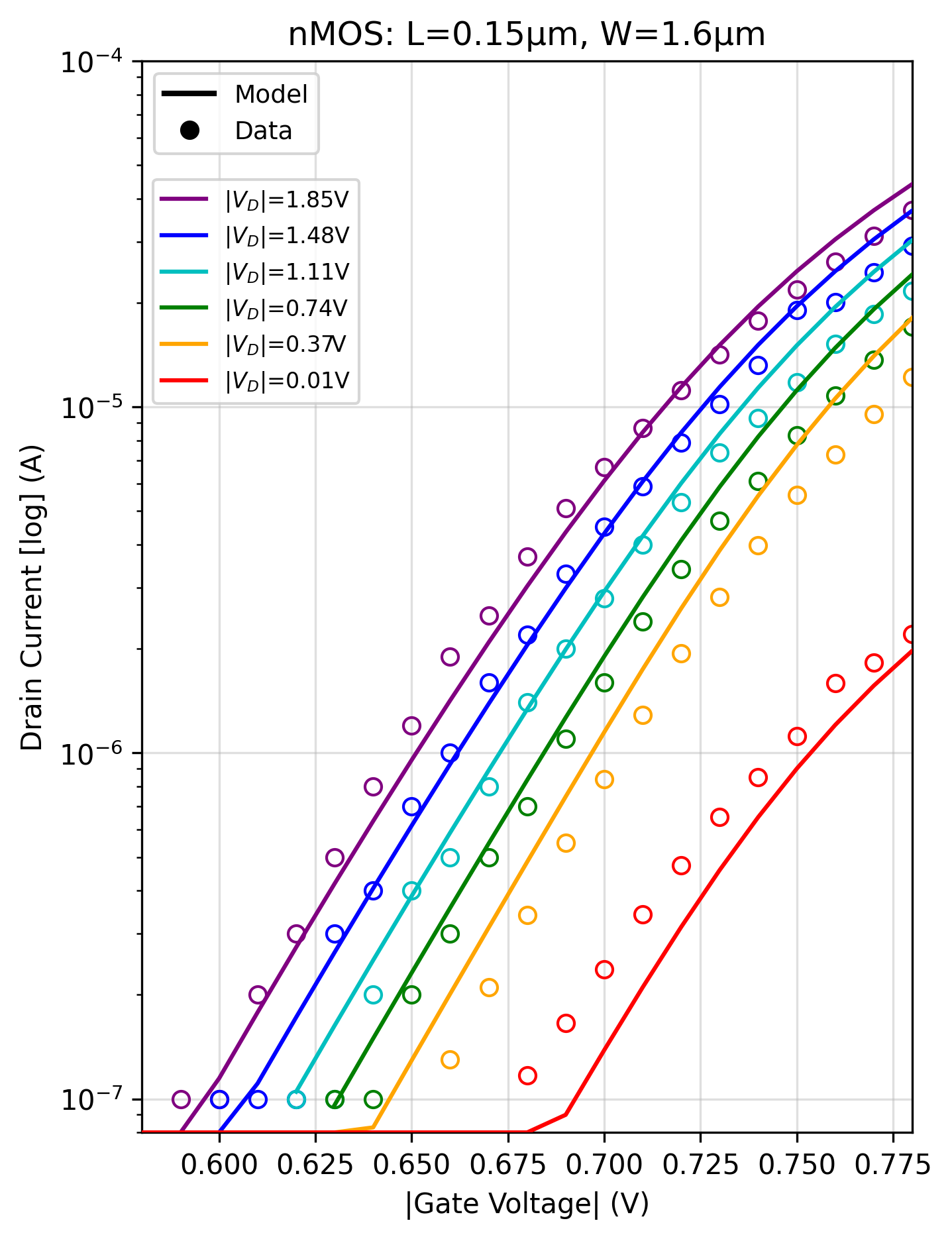} 
    \caption{}
\end{subfigure}
\hfill
\begin{subfigure}{0.22\textwidth}
    \includegraphics[width=\linewidth]{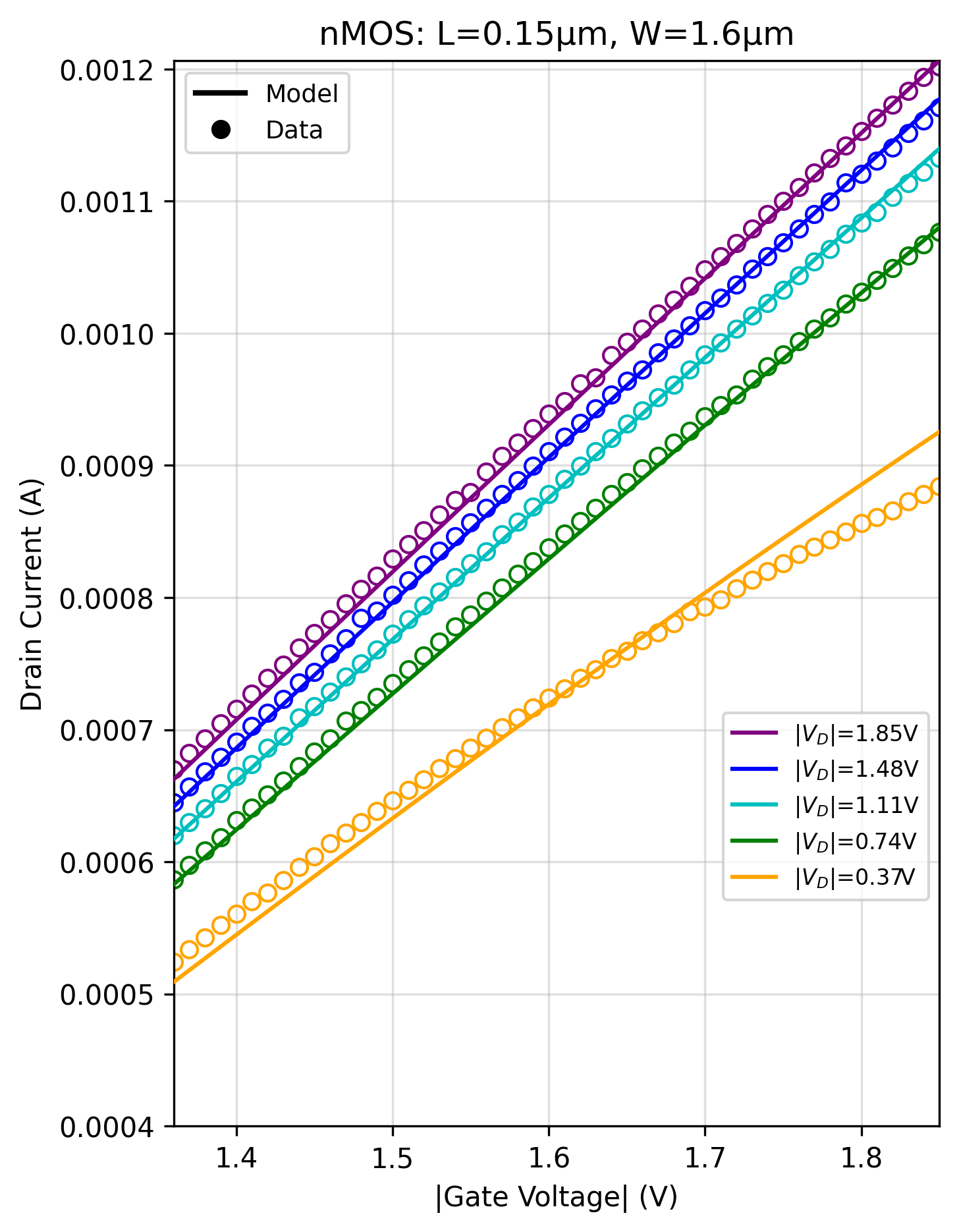} 
    \caption{}
\end{subfigure}
\hfill
\begin{subfigure}{0.22\textwidth}
    \includegraphics[width=\linewidth]{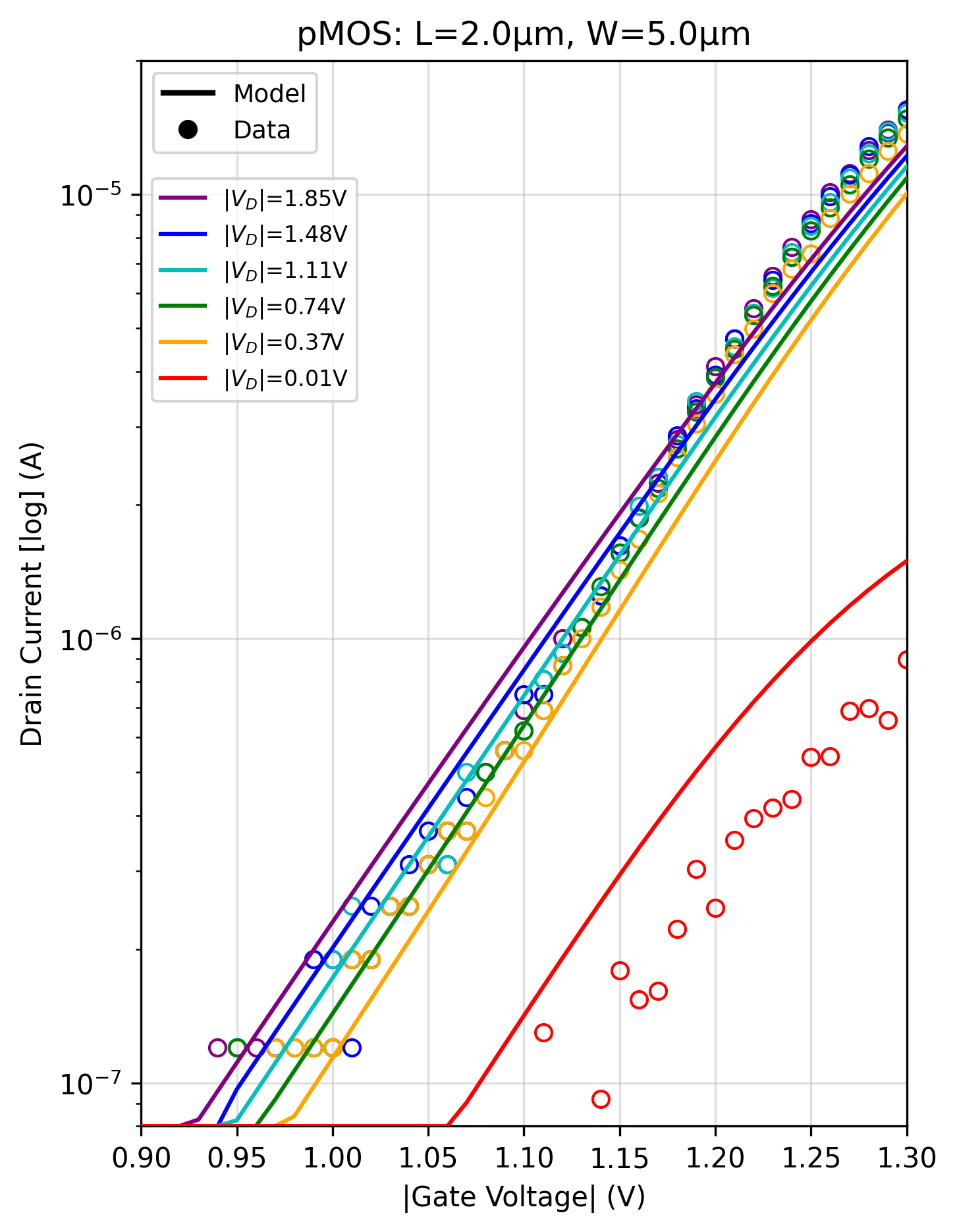} 
    \caption{}
\end{subfigure}
\hfill
\begin{subfigure}{0.24\textwidth}
    \includegraphics[width=\linewidth]{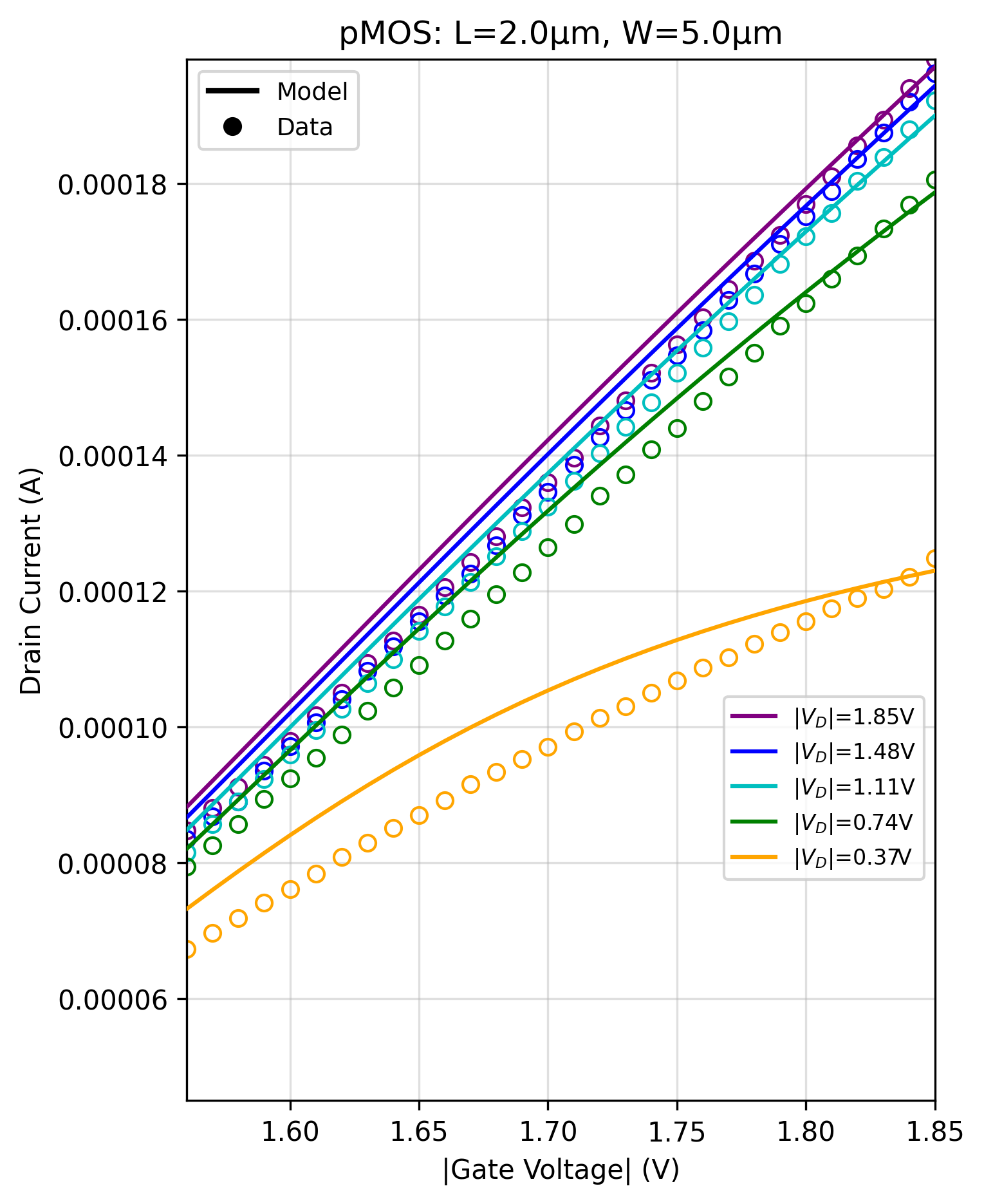} 
    \caption{}
\end{subfigure}
\hfill

\caption{Best-fit I–V characteristics for 77\,K data (hollow circles) and cryo models (solid line): (a)-(b) output and transfer characteristic curves for nMOS of length 0.15$\mu$m and width 1.6$\mu$m, (c)-(d) output and transfer characteristic curves for pMOS of length 2$\mu$m and width 5$\mu$m. The $I_D$--$V_{GS}$ plots have both linear and log scales to better observe the subthreshold region. (e)-(h) display the weak and strong inversion of the best fit transfer characteristics for a better assessment of the model fit in these regimes. The strong inversion plots focus on large $V_{GS}$ values for a better view of the data-model agreement in these regions.}
\label{fig:cryo_IV_curves}
\end{figure}

Figure~\ref{fig:cryo_IV_curves} displays the 77\,K data alongside the generated 77\,K cryogenic model
for two transistors with the best data-model agreement. The weak and strong inversion regions of both transfer characteristic curves are highlighted in (e)-(h) for a better look at the data-model agreement in those particular regions. We observe reasonable data-model agreement at high drain currents with slightly larger mismatch at lower drain current. The pMOS devices shown in Figures~\ref{fig:cryo_IV_curves}d and~\ref{fig:cryo_IV_curves}g display a visible data-model discrepancy at $|V_{DS}|=0.01\,\mathrm{V}$, due to the prioritization of the saturation fit mentioned in Section~\ref{sec:ModelAndDesignParameters}.

\subsection{Model Accuracy}
There is no single universally adopted scalar metric for quantifying how well a compact-model fit reproduces measured transistor data. Here we use a relative root-mean-square (RRMS) error to summarize the agreement between the measured and simulated drain current. For each individual bias curve, we define
\begin{equation}
    \mathrm{RRMS}_k =
    \frac{
    \sqrt{\frac{1}{N_k}\sum_{i=1}^{N_k}
    \left(I_{i,k}^{\mathrm{model}}-I_{i,k}^{\mathrm{data}}\right)^2}
    }{
    \frac{1}{N_k}\sum_{i=1}^{N_k}\left|I_{i,k}^{\mathrm{data}}\right|
    },
    \label{rrms_eq}
\end{equation}
where $I_{i,k}^{\mathrm{data}}$ and $I_{i,k}^{\mathrm{model}}$ are the measured and modeled drain-current values at the $i$th bias point of the $k$th curve, and $N_k$ is the number of data points in that curve. Here, $k$ labels either a transfer characteristic measured at fixed $V_{DS}$ while sweeping $V_{GS}$, or an output characteristic measured at fixed $V_{GS}$ while sweeping $V_{DS}$.

For each transistor, an RRMS value is first computed for every transfer and output curve individually. If a given transistor has a total of $M$ such curves, we then report the mean RRMS
\begin{equation}
    \overline{\mathrm{RRMS}}=
    \frac{1}{M}\sum_{k=1}^{M}\mathrm{RRMS}_k,
\end{equation}
together with its population standard deviation
\begin{equation}
    \sigma_{\mathrm{RRMS}}=
    \sqrt{\frac{1}{M}\sum_{k=1}^{M}
    \left(\mathrm{RRMS}_k-\overline{\mathrm{RRMS}}\right)^2 }.
\end{equation}
$\overline{\mathrm{RRMS}}$ provides a single measure of the overall goodness-of-fit for a transistor model, while the standard deviation indicates how consistently that fit quality is maintained across different bias conditions. Smaller values of $\overline{\mathrm{RRMS}}$ indicate better overall agreement, and smaller values of $\sigma_{\mathrm{RRMS}}$ indicate more uniform model accuracy across the set of measured curves.

\begin{figure}[htbp]
\centering

\begin{subfigure}{0.48\textwidth}
    \centering
    {%
    \includegraphics[width=\linewidth]{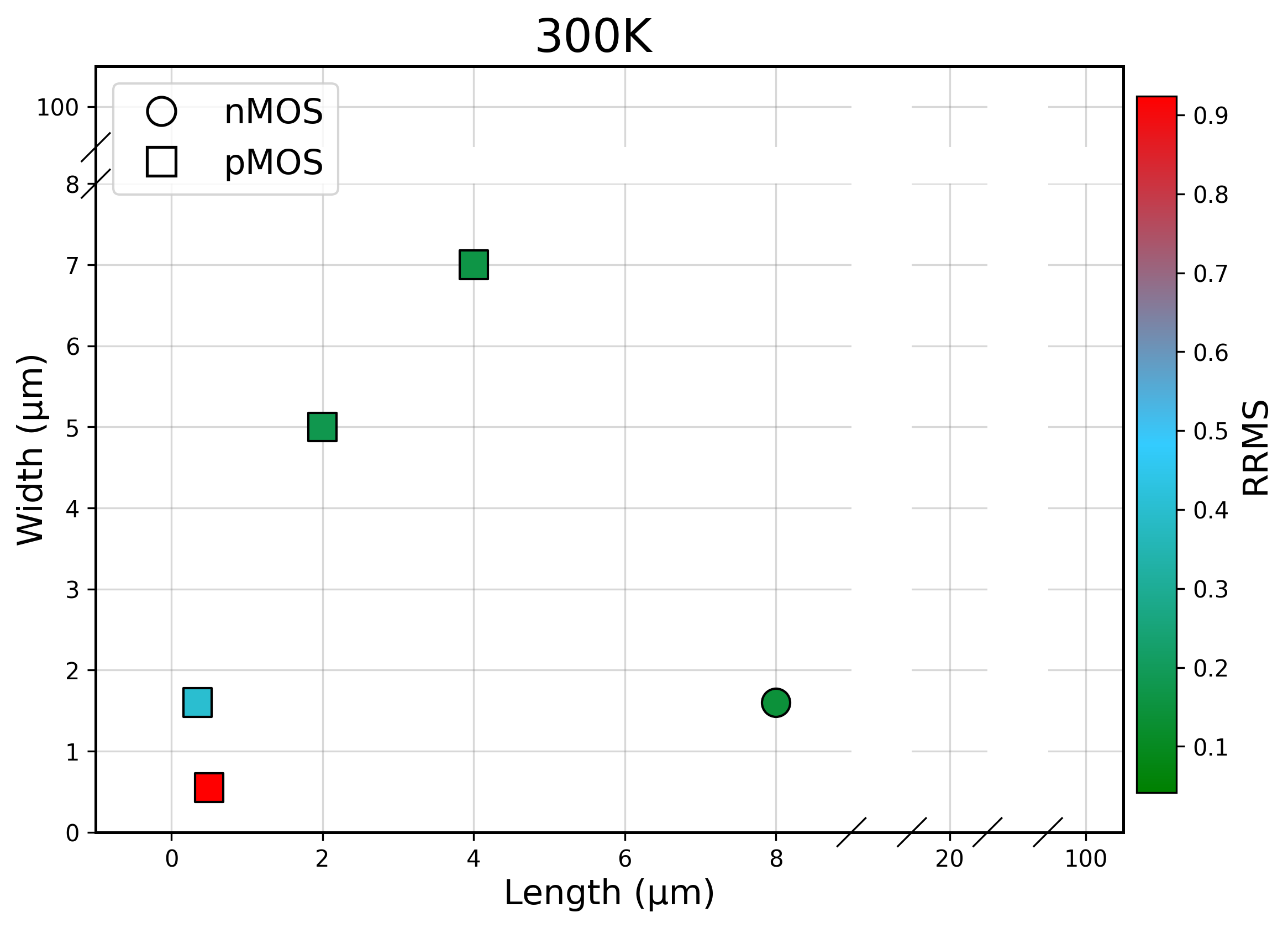}%
}
    \caption{}
    \label{fig:rrms_300K}
\end{subfigure}
\hfill
\begin{subfigure}{0.48\textwidth}
    \centering
        {%
    \includegraphics[width=\linewidth]{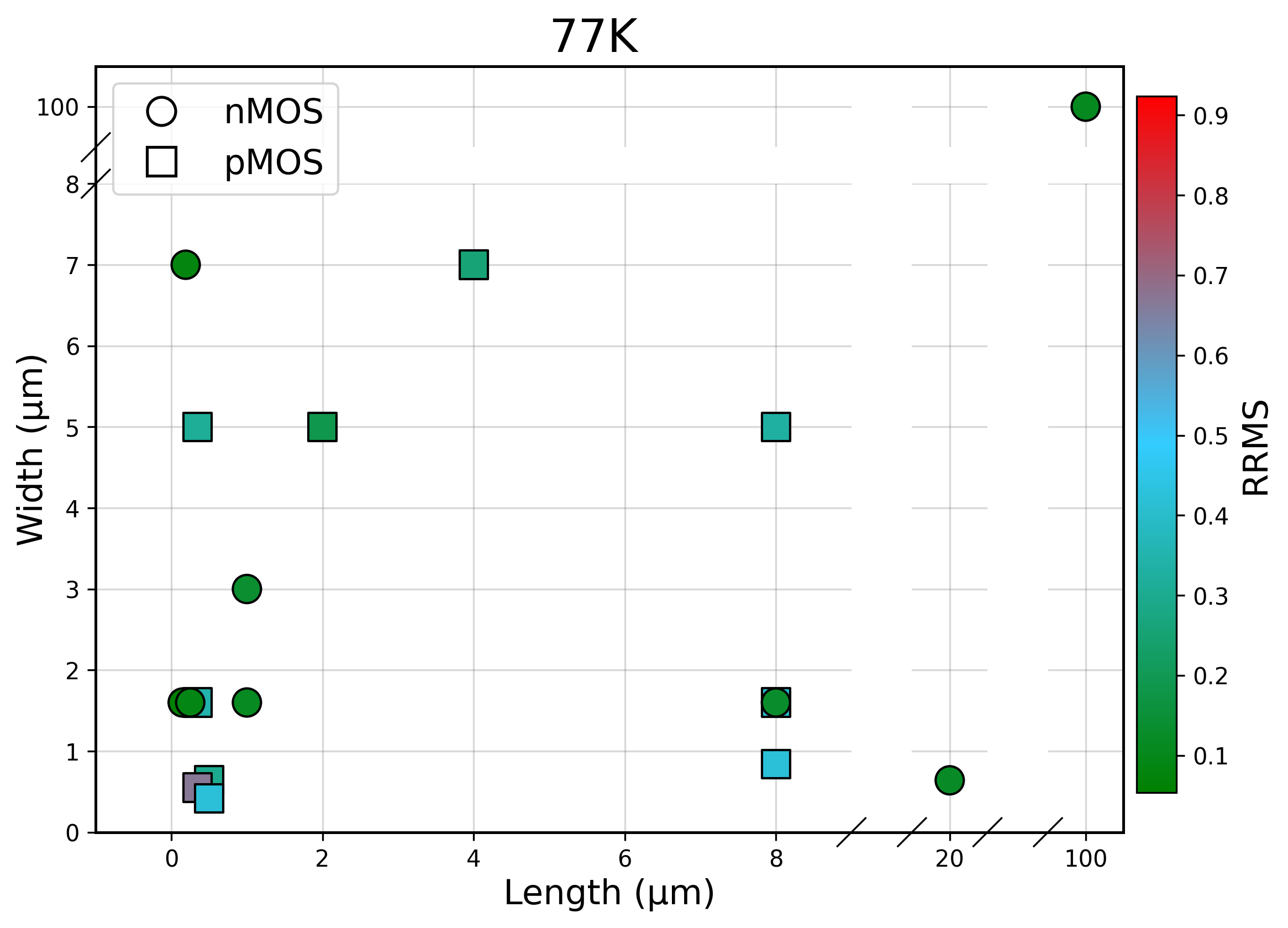}%
    }
    \caption{}
    \label{fig:rrms_77K}
\end{subfigure}

\caption{Data-model agreement (RRMS, see Equation \ref{rrms_eq}) at (a) 300\,K and (b) 77\,K. Both plots use the same minimum and maximum RRMS for direct comparison.}
\label{fig:model_fit_quality}

\end{figure}

Figure \ref{fig:model_fit_quality} displays the $\overline{\mathrm{RRMS}}$ for each transistor of different lengths and widths tested at room temperature and 77\,K in graphical form. A complete overview of the mean root mean square error ($\overline{\mathrm{RMSE}}$), $\overline{\mathrm{RRMS}}$, and $\sigma_{\mathrm{RRMS}}$ values for each transistor at 300\,K and 77\,K are contained in Appendix \ref{appendix:error}. The RMSE of one I–V curve is the same as Equation \ref{rrms_eq} without normalization. The reported $\overline{\mathrm{RMSE}}$ is averaged over all curves for each transistor.

For each transistor, the mean RRMS is the arithmetic mean of the separately normalized RRMS values of its individual transfer and
output curves, as defined in Eqs.~(5.1)--(5.3). It summarizes agreement over the measured data set but does not establish uniform accuracy at every bias. In particular, the systematic pMOS model--data discrepancy at $|V_{DS}|=0.01~\mathrm{V}$ represents
a bias-dependent limitation that must be considered alongside the mean RRMS and its standard deviation across curves, $\sigma_{\mathrm{RRMS}}$. Model accuracy therefore depends on
device bias conditions.

\newpage
\section{Conclusion}\label{sec:conclusion}
This work presents the first systematic DC characterization and open-source BSIM4-based isothermal model of SKY130 LVT MOSFETs at 77\,K. Using a per-bin extraction strategy applied to measured 77\,K output and transfer characteristics, we developed eighteen distinct cryogenic models at 77\,K spanning the measured nMOS and pMOS geometries. The resulting models improve agreement with measured 77\,K device behavior relative to the room-temperature PDK, with average errors on the order of $20\%$ in relative RMS. Additionally, the relative RMS errors in the 77\,K models are comparable, and in some cases, better than, the room temperature data-model agreement. However, a systematic model--data discrepancy persists in the transfer characteristics at the lowest measured drain bias
($|V_{DS}|=0.01~\mathrm{V}$). This discrepancy is observed in both nMOS and pMOS devices but is more pronounced in pMOS. Its physical
origin remains unresolved, and the reported parameterization
should therefore be used with this limitation in mind. These results establish SKY130 as a practical open-source node for 77\,K cryogenic circuit design in high-energy physics and related low-temperature instrumentation applications. Our developed models are NGSpice-compatible and publicly available on GitHub \cite{github}. The present work is limited to DC isothermal modeling and does not yet include noise, capacitance, and has not yet experimentally validated mismatch or process corner variations. Future work will extend the model across additional cryogenic temperatures and broader geometry coverage toward a temperature- and geometry-dependent cryogenic PDK. More broadly, the 77\,K DC model developed here provides a foundation for future cryogenic analog circuit design in this open-source technology node.

\acknowledgments
The authors would like to thank various engineers in the microelectronics department at FNAL for their guidance and assistance on this project:  Albert Dyer for help operating the cryo-cooler, and Louis Dal Monte and Pamela Klabbers for PCB design. The authors would also like to extend gratitude to Andy Pender from Synopsys for assistance with the modeling software, Mystic\texttrademark. This material is also based upon work supported by U.S. Department of Energy, Office of Science, Office of High Energy Physics under Award Number DE-SC0022296 and DE-SC00253485 as well as support from the University of Texas at Arlington's Center for Advanced Detector Technologies. JBRB acknowledges the support of the Gordon and Betty Moore Foundation (Grant No. GBMF11565).

\newpage
\newpage
\appendix
\section{Error Tables}
\label{appendix:error}
\begin{table}[htbp]
    \centering
    \renewcommand{\arraystretch}{1.2}
    \begin{tabularx}{\textwidth}{|c|c| c| c| C| C|}
    \hline
    Transistor Type & Length ($\mu$m) & Width ($\mu$m) & $\overline{\mathrm{RMSE}}$ ($\mu$A) & $\overline{\mathrm{RRMS}}$ & $\sigma_{\mathrm{RRMS}}$  \\
    \hline 
    nMOS & 8 & 1.6 & 1.43 & 0.142 & 0.105 \\
    \hline
    pMOS & 0.35 & 1.6 & 59.5 & 0.377 & 0.154 \\
     & 0.5 & 0.55 & 8.87 & 0.924 & 0.650 \\
     & 2 & 5 & 2.49 & 0.177 & 0.240 \\ 
     & 4 & 7 & 1.38 & 0.164 & 0.253 \\
    \hline
    \end{tabularx}
    \caption{Reported errors for 300\,K models for select transistor sizes.}
\end{table}
\begin{table}[htbp]
    \centering
    \renewcommand{\arraystretch}{1.2}
    \begin{tabularx}{\textwidth}{|c|c| c| c| C| C|}
    \hline
    Transistor Type & Length ($\mu$m) & Width ($\mu$m) & $\overline{\mathrm{RMSE}}$ ($\mu$A) & $\overline{\mathrm{RRMS}}$ & $\sigma_{\mathrm{RRMS}}$  \\
    \hline 
    nMOS & 0.15 & 1.6 & 11.6 & 0.054 & 0.064 \\
    & 0.19 & 7 & 70.0 & 0.084 & 0.118 \\
    & 0.25 & 1.6 & 16.9 & 0.088 & 0.097 \\
    & 1 & 1.6 & 11.6 & 0.112 & 0.144 \\
    & 1 & 3 & 41.2 & 0.138 & 0.166 \\
    & 8 & 1.6 & 3.02 & 0.131 & 0.140 \\
    & 20 & 0.64 & 0.702 & 0.119 & 0.078 \\
    & 100 & 100 & 20.6 & 0.111 & 0.111 \\
    \hline
    pMOS & 0.35 & 0.55 & 1.45 & 0.667 & 1.48 \\
    & 0.35 & 1.6 & 9.25 & 0.349 & 0.741 \\
    & 0.35 & 5 & 32.0 & 0.312 & 0.463 \\
    & 0.5 & 0.42 & 2.24 & 0.422 & 0.738 \\
    & 0.5 & 0.64 & 1.35 & 0.300 & 0.588 \\
    & 2 & 5 & 2.73 & 0.187 & 0.364 \\
    & 4 & 7 & 4.92 & 0.255 & 0.376 \\
    & 8 & 0.84 & 0.255 & 0.421 & 0.763 \\
    & 8 & 1.6 & 0.424 & 0.410 & 0.796 \\
    & 8 & 5 & 1.56 & 0.330 & 0.561 \\
    \hline
    \end{tabularx}
    \caption{Reported errors for developed 77\,K models for select transistor sizes.}
\end{table}
\newpage
\bibliographystyle{JHEP}
\bibliography{skywater} 

\end{document}